\pdfoutput=1
\documentclass[aps,prd,twocolumn,epsf,superscriptaddress,,tightenlines,nofootinbib]{revtex4}

\usepackage{graphicx}
\usepackage{amsmath,amssymb,latexsym}
\usepackage{bm}
\usepackage{color}

\def\lbldef#1#2{\expandafter\gdef\csname #1\endcsname {#2}}

\def\href#1#2{#2}

\newcommand{\ber}{\begin{eqnarray}}
\newcommand{\eer}{\end{eqnarray}}

\newcommand{\beqar}{\begin{eqnarray}}

\newcommand{\eeqar}{\end{eqnarray}}

\newcommand{\dsl}
  {\kern.06em\hbox{\raise.15ex\hbox{$/$}\kern-.56em\hbox{$\partial$}}}

\newcommand{\eeqarr}{\end{eqnarray}}
\newcommand{\ZZ}{{\rm \kern 0.275em Z \kern -0.92em Z}\;}

\begin{document}

\title{Summary Report of JEM-EUSO Workshop at KICP in Chicago}

\author{James H. Adams Jr.}
\affiliation{CSPAR, University of Alabama in Huntsville, Huntsville,
  AL 35805, USA}

\author{Luis A.~Anchordoqui}
\affiliation{Department of Physics,
University of Wisconsin-Milwaukee,
 Milwaukee, WI 53201, USA}

\author{Mario Bertaina}
\affiliation{Dipartamento di Fisica Generale, Universit\'a di Torino, Torino, Italy}

\author{Mark J. Christl}
\affiliation{Marshall Space Flight Center, Huntsville, AL 35812, USA}

\author{Valerie \nolinebreak Connaughton} 
\affiliation{CSPAR, University of Alabama in Huntsville, Huntsville,
  AL 35805, USA}

\author{Steven~E. \nolinebreak Csorna}
\affiliation{Department of Physics and Astronomy,
Vanderbilt University, Nashville TN 37235, USA}

\author{Toshikazu Ebisuzaki}
\affiliation{RIKEN Advanced Science Institute, 2-1 Hirosawa, Wako 351-0198, Japan}

\author{Gustavo Medina-Tanco}
\affiliation{Instituto de Ciencias Nucleares, UNAM, Ciudad Universitaria, M\'exico D.F. 04510, M\'exico}

\author{Angela V. Olinto} \affiliation{Department of Astronomy and
  Astrophysics, University of Chicago, Chicago, IL 60637, USA}

\author{Thomas Paul}
\affiliation{Department of Physics,
University of Wisconsin-Milwaukee,
 Milwaukee, WI 53201, USA}
\affiliation{Department of Physics,
Northeastern University, Boston, MA 02115, USA}

\author{Piergiorgio  \nolinebreak Picozza}
\affiliation{INFN, Sezione di Rome Tor Vergata, I-00133 Rome, Italy}

\author{Andrea Santangelo}
\affiliation{Institut f\"ur Astronomie und Astrophysik, Eberhard-Karls 
Universit\"at T\"ubingen, T\"ubingen, Germany}

\author{Kenji Shinozaki}
\affiliation{RIKEN Advanced Science Institute, 2-1 Hirosawa, Wako 351-0198, Japan}

\author{Thomas J. Weiler}
\affiliation{Department of Physics and Astronomy,
Vanderbilt University, Nashville TN 37235, USA}

\author{Lawrence Wiencke} \affiliation{Department of Physics, Colorado
  School of Mines, Golden, CO 80401, USA}

\date{March 2012}
\begin{abstract}
  \noindent This document contains a summary of the workshop which
took place on 22--24 February 2012 at  the Kavli Institute of Cosmological Physics at the University of Chicago. The goal of the workshop was to discuss the physics reach of the JEM-EUSO mission and how best to implement a global ground based calibration system for the instrument to realize the physics goal of unveiling the origin of the highest energy cosmic rays.  
\end{abstract}


\maketitle

\section{Introduction}

Ultra high energy cosmic rays (UHECRs) are one of the most enigmatic
phenomena in the universe. Despite the fact that the existence of
particles with energies reaching $10^{20}$~eV has been known for over
50 years~\cite{Linsley:1963km}, their origin continues to be an intriguing
puzzle~\cite{Hillas:1985is,Nagano:2000ve,Torres:2004hk,Beatty:2009zz,Kotera:2011cp,LetessierSelvon:2011dy}.

Soon after the microwave echo of the big bang was discovered,
Greisen~\cite{Greisen:1966jv}, Zatsepin, and
Kuzmin~\cite{Zatsepin:1966jv} (GZK) noted that the relic photons make
the universe opaque to cosmic rays (CRs) of sufficiently high energy.  This
occurs, for example, for protons with energies beyond the photopion
production threshold,
\begin{eqnarray}
E_{p\gamma_{\rm CMB}}^{\rm th} & = & \frac{m_\pi \, (m_p + m_\pi/2)}{\omega_{\rm CMB}} \nonumber \\
& \approx & 6.8 \times 10^{19}\,
\left(\frac{\omega_{\rm CMB}}{10^{-3}~{\rm eV}}\right)^{-1}~{\rm eV}\,,
\label{1}
\end{eqnarray}
where $m_p$ ($m_\pi$) denotes the proton (pion) mass and $\omega_{\rm
  CMB} \sim 10^{-3}$~eV is a typical photon energy of the cosmic
microwave background (CMB). After pion production, the proton (or
perhaps, instead, a neutron) emerges with at least 50\% of the
incoming energy. This implies that the nucleon energy changes by an
$e$-folding after a propagation distance~$\lesssim
(\sigma_{p\gamma}\,n_\gamma\,y_\pi)^{-1} \sim
15$~Mpc~\cite{Stecker:1968uc,Berezinsky:1988wi,Aharonian:1994nn,Anchordoqui:1996ru}. Here,
$n_\gamma \approx 410$~cm$^{-3}$ is the number density of the CMB
photons, $\sigma_{p \gamma} > 0.1$~mb is the photopion production
cross section, and $y_\pi$ is the average energy fraction (in the
laboratory system) lost by a nucleon per interaction. For heavy
nuclei, the giant dipole resonance can be excited at similar total
energies and hence, for example, iron nuclei do not survive
fragmentation over comparable
distances~\cite{Stecker:fw,Puget:nz,Anchordoqui:1997rn,Epele:1998ia,Stecker:1998ib,
  Khan:2004nd,Allard:2005ha,Hooper:2008pm}. Additionally, the survival
probability for extremely high energy ($\approx 10^{20}$~eV)
$\gamma$-rays (propagating on magnetic fields~$\gg 10^{-11}$~G) to a
distance $d$, \mbox{$P(>d) \approx \exp[-d/6.6~{\rm Mpc}]$}, becomes
less than $10^{-4}$ after traversing a distance of
50~Mpc~\cite{Elbert:1994zv}. This implies that the GZK sphere
represents a small fraction of the size of the universe.\footnote{The
  sphere within which a source has to lie in order to provide us with
  UHECRs if the primaries are subject to the GZK phenomenon.}
Consequently, if the CR sources are universal in origin, the energy
spectrum should have a greatly reduced intensity
beyond $E \sim E_{\rm GZK} \equiv 6  \times 10^{19}$~eV,
a phenomenon known as the GZK suppression.

The CR spectrum spans over roughly 11 decades of energy.  Its shape is
remarkably featureless, with little deviation from a constant power
law ($J \propto E^{-\gamma}$, with $\gamma \approx 3$) across this
large energy range. In 2007, the HiRes Collaboration reported a
suppression of the CR flux above $E = [5.6 \pm 0.5 ({\rm stat}) \pm
0.9 ({\rm syst})] \times 10^{19}~{\rm eV}$, with 5.3$\sigma$
significance~\cite{Abbasi:2007sv}. The spectral index of the flux
steepens from $2.81\pm 0.03$ to $5.1 \pm 0.7$. The discovery of the
suppression has been confirmed by the Pierre Auger Collaboration,
measuring $\gamma = 2.69 \pm 0.2 ({\rm stat}) \pm 0.06 ({\rm syst})$
and $\gamma = 4.2 \pm 0.4 ({\rm stat}) \pm 0.06 ({\rm syst})$ below
and above $E = 4.0 \times 10^{19}~{\rm eV}$, respectively (the
systematic uncertainty in the energy determination is estimated as
22\%)~\cite{Abraham:2008ru}. 
\begin{figure}
\centering
\includegraphics[width=0.5\textwidth]{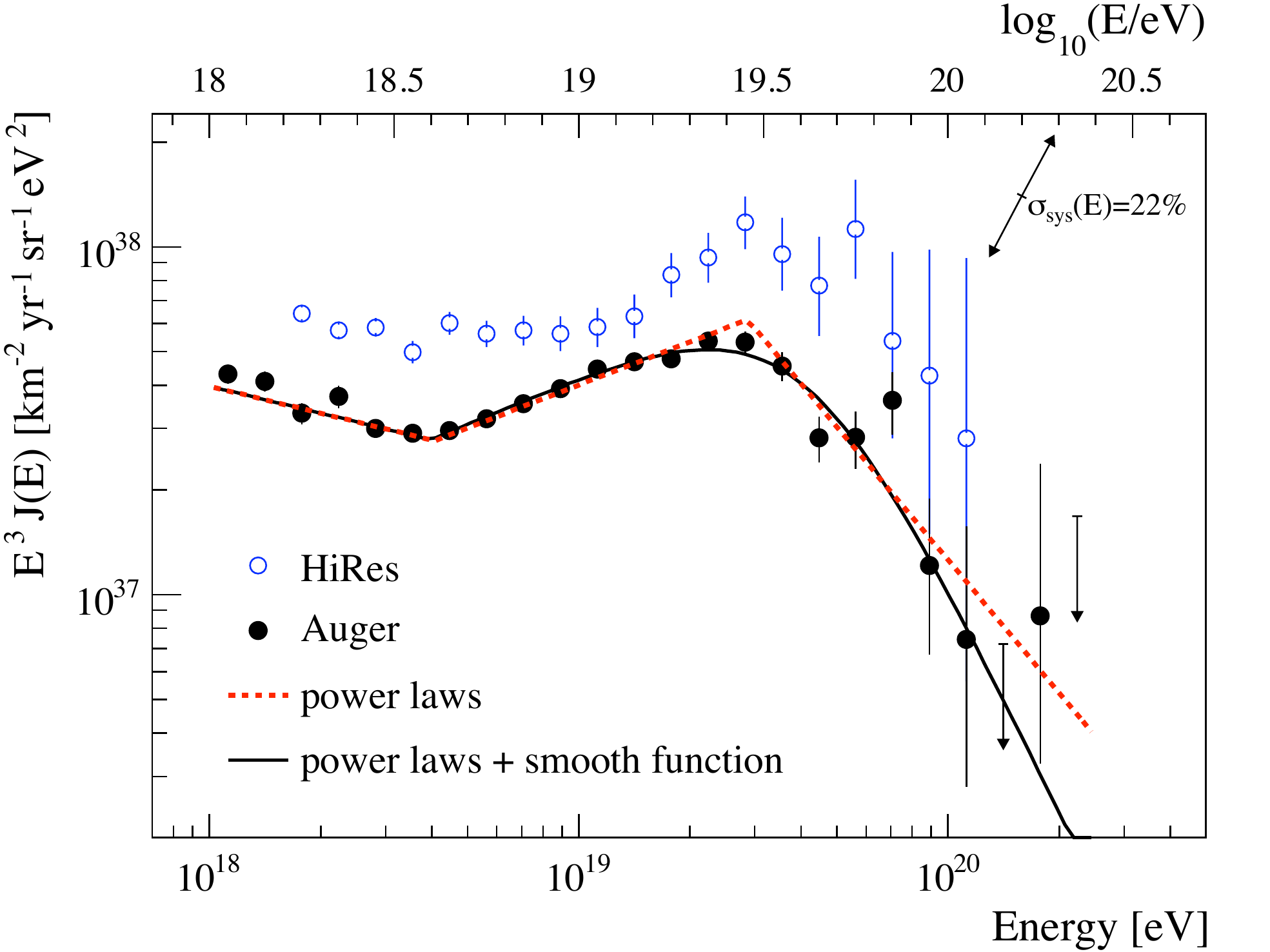}
\caption{Combined spectrum from Auger (hybrid and SD events) and the stereo spectrum HiRes.  The Auger systematic uncertainty of  the flux scaled by $E^3$, due to the uncertainty of the energy scale of 22\%, is  indicated by arrows. The results of the two experiments are consistent within systematic uncertainties. From Ref.~\cite{Abraham:2010mj}.}
\label{fig:cal}
\end{figure}

\begin{figure*}[htbp]
\begin{center}
\includegraphics[width=0.9\textwidth]{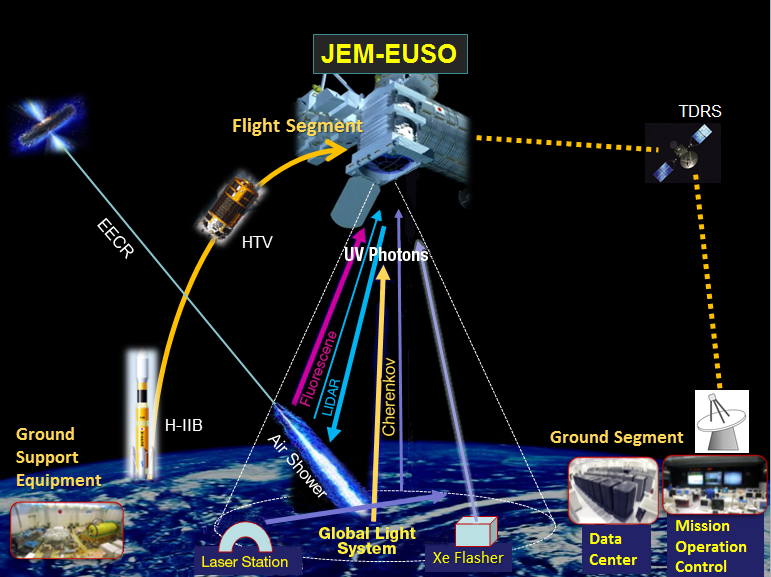}
\caption{JEM-EUSO overview and concept of mission operations.} 
\label{CoM}
\end{center}
\end{figure*}

In 2010, an updated Auger measurement of the energy spectrum was
published~\cite{Abraham:2010mj}, corresponding to a surface array
exposure of $12,790~{\rm km}^2~{\rm sr~yr}$.  This measurement,
combining both hybrid and surface detector (SD)-only events, is shown
in Fig.~\ref{fig:cal}.  The so-called ``ankle'' feature and the flux
suppression are clearly visible.  A broken power law fit to the
spectrum shows that the break corresponding to the ankle is located at
$\log_{10}(E/{\rm eV}) = 18.61 \pm 0.01$ with $\gamma = 3.26 \pm 0.04$
before the break and $\gamma = 2.59 \pm 0.02$ after it.  The break
corresponding to the suppression is located at $\log_{10}(E/{\rm eV})
= 19.46 \pm 0.03$.  Compared to a power law extrapolation, the
significance of the suppression is greater than $20\sigma$. 

The extreme energy  ($E \ge E_{\rm GZK}$) cosmic ray (EECR)
flux is consequently exceptionally low, of the order of 1~{\em particle/km$^2$/sr/century}. 
At the high end of the spectrum, $E > 10^{20}~{\rm eV}$, it reduces to about 1~{\em
  particle/km$^2$/sr/millennium!} This challenging extreme energy
region is the scope of the Extreme Universe Space Observatory (EUSO)
attached to the Japanese Experiment Module (JEM) on board the
International Space Station (ISS)~\cite{Takahashi:2009zzc}.

Currently the leading observatories of EECRs are ground based
observatories that cover vast areas with particle detectors overlooked
by fluorescence telescopes. The largest is the Pierre Auger
Observatory in Argentina with a surface detector array of 1600 water
Cherenkov tanks covering 3,000 km$^2$ which accumulates annually about
$6 \times 10^3$ km$^2$~sr~yr of exposure~\cite{Abraham:2010zz}. The
more recently constructed Telescope Array (TA) covers 700 km$^2$ with
507 scintillator detectors~\cite{Tsunesada:2011mp}, which should
accumulate annually about $1.4 \times 10^3$~km$^2$~sr~yr of exposure.

Although extremely large, current observatories are not large enough
to study EECRs with the necessary statistics. JEM-EUSO will observe
over $4 \times 10^4$ km$^2$~sr~yr at $E_{\rm GZK}$ and reaches 100\%
exposure around $10^{20}$~eV where it observes $6 \times 10^4$
km$^2$~sr~yr annually~\cite{Shinozaki,Adams:2012tt}, a factor of 10 above
Auger. With a launch date planned for 2017, JEM-EUSO will accumulate
$3 \times 10^5$ km$^2$~sr~yr by 2022 in Nadir mode or it can reach the
target of $10^6$ km$^2$~sr~yr during the same period in a Tilted mode.

A significant increase in exposure is needed to identify the first
sources of UHECRs which should make a clear imprint in the sky at
extreme energies (above $E_{\rm GZK}$). Anisotropies in the
distribution of arrival directions of EECRs are a consequence of the
GZK suppression. The GZK effect limits the distance from which sources
of extreme energy hadrons can contribute to the flux on Earth to below
$\sim 100$ Mpc~\cite{Harari:2006uy,Allard:2005cx}. The distribution of
matter within this distance from Earth is anisotropic, therefore the
sky distribution of EECRs should also be anisotropic, modulo the
effect of extragalactic magnetic fields. For proton primaries, the
anisotropy should be apparent around $E_{\rm GZK}$, while for higher
charged heavier nuclei the anisotropy pattern should become apparent
at higher energies, as discussed in Sec.~\ref{sec:ani}  below.
 
JEM-EUSO will be the first mission to observe UHECRs from space. This pioneering mission may be the first to clearly identify a source of UHECRs by observing a significantly larger number of EECRs. The source signature will be observed both through the details of the anisotropic distribution of arrival directions as well as the detailed shape of the spectrum above $E_{\rm GZK}$. Both observations will answer the question of whether the observed spectral suppression is the GZK effect or the maximum energy of UHECR accelerators. Once a clear source is identified, a space based program can be developed to observe the ten million EECR particles that reach the Earth's atmosphere every year.

\section{JEM-EUSO Mission}
\label{sec:jem-euso}

JEM-EUSO is an innovative {\em pathfinder} space mission that will
exploit the Earth's atmosphere as a detector of cosmic ray showers.
The remote-sensing space instrument would orbit the Earth every $\sim
90$ minutes on the ISS at an altitude $h_{\rm ISS} = 350 - 400~{\rm
  km}$. The EUSO instrument is a  2.5 meter  telescope
with a high-speed ultraviolet (UV) camera  (2.5 $\mu$s
  resolution) that will observe the nighttime atmosphere below the
ISS with a 60$^{\circ}$ field of view
  (FOV)~\cite{Kajino,Adams:2012tt}.  It will monitor an area larger than 1.3
  $\times 10^5$~km$^2$ recording  video clips of fast UV flashes by
sensing the fluorescence light produced through charged particle
interactions (see Fig.~\ref{CoM}).  This fluorescence is detectable in
the $300 - 400$~nm range.  JEM-EUSO will have about 3
  $\times 10^5$ pixels each with less than 3 mm that cover about 560 m
  on the ground.

  When the incident cosmic ray interacts with the atomic nuclei of air
  molecules it produces extensive air showers (EAS) which spread out
  over large areas (see
  e.g.~\cite{Anchordoqui:2004xb,Matthews:2005sd}). JEM-EUSO will image
  light from isotropic nitrogen fluorescence excited by the EAS, and
  forward-beamed Cherenkov radiation reflected from the Earth's
  surface or dense cloud tops. The video recorded by JEM-EUSO can
  capture the moving track of the fluorescent UV photons and
  reproduces the temporal development of EAS. The instrument will be
  calibrated using ground-based lasers and Xe flashers. Description of
  such a calibration procedure is postponed to Sec.~\ref{sec:GLS}.

  Fluorescence observations of the longitudinal development of EAS can
  give a good estimate of the energy, since the fluorescence light is
  (in principle) proportional to the electromagnetic (electrons,
  positrons, and photons) component of the shower which gives a
  calorimetric estimate of the shower energy.  The longitudinal
  development also has a well defined maximum, usually referred to as
  $X_{\rm max}$, which increases with primary energy as more cascade
  generations are required to degrade the energy of the secondary
  particles. The nature of the primary is harder to determine due to
  large fluctuations of the first interaction point, but the average
  depth of shower maximum, $\left<X_{\rm max}\right> \propto \ln(E/A)$
  where $A$ is the mass of the primary cosmic ray nucleus of energy
  $E$, can be related to the primary mass through a comparison with
  hadronic interaction models~\cite{Linsley:gh}.  A number of
  alternative methods for mass determinations have also been used
  based both on the longitudinal development of the fluorescence
  signal and the lateral distributions of particles on the ground
  (see, e.g.,~\cite{Kampert:2012mx} for a recent review).

Hadronic interaction models used in describing EAS have to be extrapolated
to interactions with center-of-mass energies above 100 TeV, which is
well beyond the reach of collider experiments; 14 TeV is the reach
of the Large Hadron Collider (LHC). Current models seem to describe recent
LHC data well~\cite{dEnterria:2011jc} and indicate interesting
composition trends at the highest energies~\cite{Kampert:2012mx}. The
identification of a clear source of EECRs in the sky will allow for an
astrophysical determination of the primary charge of the EECRs which
in turn can be used to test the accuracy of hadronic interaction
models significantly enlarging the energy reach of experimental tests
of particle interactions.

JEM-EUSO will have a reasonable energy resolution, around 30\%, while
the resolution in $X_{\rm max}$ will be limited to 120 g/cm$^2$ which
does not allow a separation of different types of nuclei, but can be
used to distinguish neutrino and photon events from hadronic
events~\cite{Supanitsky:2011jp,Gus,Adams:2012tt}. Many neutrino events
have $X_{\rm max}$ deeper than $2,000~{\rm g} \,{\rm cm}^{-2},$ while
most hadronic events have $X_{\rm max}$ shallower than $2,000~{\rm
  g}\, {\rm cm}^{-2}$.

In clear air most of the light seen by JEM-EUSO as the shower develops
is from nitrogen fluorescence. Nitrogen is excited by the energetic
particles in the EAS (almost all of them are electrons). The nitrogen
can de-excite in two ways, by collision and by photon emission. At sea
level collisional de-excitation dominates. This is because the mean
time for de-excitation by photon emission is much longer than the mean
time between collisions. As a function of increasing altitude, an
increasing fraction of the excited nitrogen de-excites by photon
emission because the mean time between collisions decreases with the
decreasing atmospheric pressure. Of course, the rate of nitrogen
excitation also decreases with atmospheric pressure. The two effects
are offsetting, with the result that the fluorescence yield is
practically independent of altitude through most of the troposphere.

At higher altitudes, as collisional de-excitation becomes less
important, the fluorescence yield (photons per electron per meter
traveled in the atmosphere) decreases. This decrease affects the 337
and 357~nm lines first. Because the 391~nm line has a longer
fluorescent decay time, the fluorescence yield starts to decrease at a
much higher altitude. In most of the troposphere, the fluorescence
strength in these three lines is practically equal, but in the upper
atmosphere the 391~nm line dominates.

The mechanism for nitrogen fluorescence in EAS has two
consequences. First, the fluorescence signal is not a single
calorimetric measure of the deposited energy. The EAS must be modeled
taking into account the competition between collisional and
fluorescent de-excitation (among other effects) to determine the
energy of the cosmic ray. JEM-EUSO will not receive a fluorescent
signal that is simply proportional to the energy deposited by the EAS.

The second consequence is that the fluorescent signal seen by JEM-EUSO
depends on the zenith angle of the EAS. To understand this, consider
an EAS near $X_{\rm max}$  where the number of particles in the shower
remains nearly constant. The signal per meter the EAS travels is
proportional to the number of particles in the shower. It is easy to
see that an inclined shower will reach $X_{\rm max}$  at a higher
altitude where the atmospheric pressure is lower. At such an altitude the
shower maximum will be stretched out over a greater distance since its
length depends on the grammage traversed by the EAS, not
distance. Since the fluorescent signal is proportional to the number
of charged particles, the longer distance near shower maximum results
in a larger fluorescent signal.

The JEM-EUSO telescope concept is a fast, highly-pixelized, large aperture and
wide Field-of-View ($\pm 30^\circ$ FOV) digital
camera~\cite{Kajino:2011zz,Kajino,Kawasaki:2011zz}. In the baseline design
the total number of pixels is $\sim 3 \times 10^5$. The angular
resolution of each pixel is about $0.07^\circ$. For $h_{\rm ISS} =
400~{\rm km}$, this resolution corresponds to $\sim 0.5~{\rm km}$ on
the Earth's surface.  For each pixel, data is acquired with a time
resolution, a.k.a. gate time unit (GTU), of $2.5~\mu{\rm s}$. These time-sliced
images allow determination of the energy and incoming direction of the
primary particle~\cite{Ebisuzaki:2010zz}.

\begin{figure}[htbp]
\begin{center}
\includegraphics[width=0.4 \textwidth]{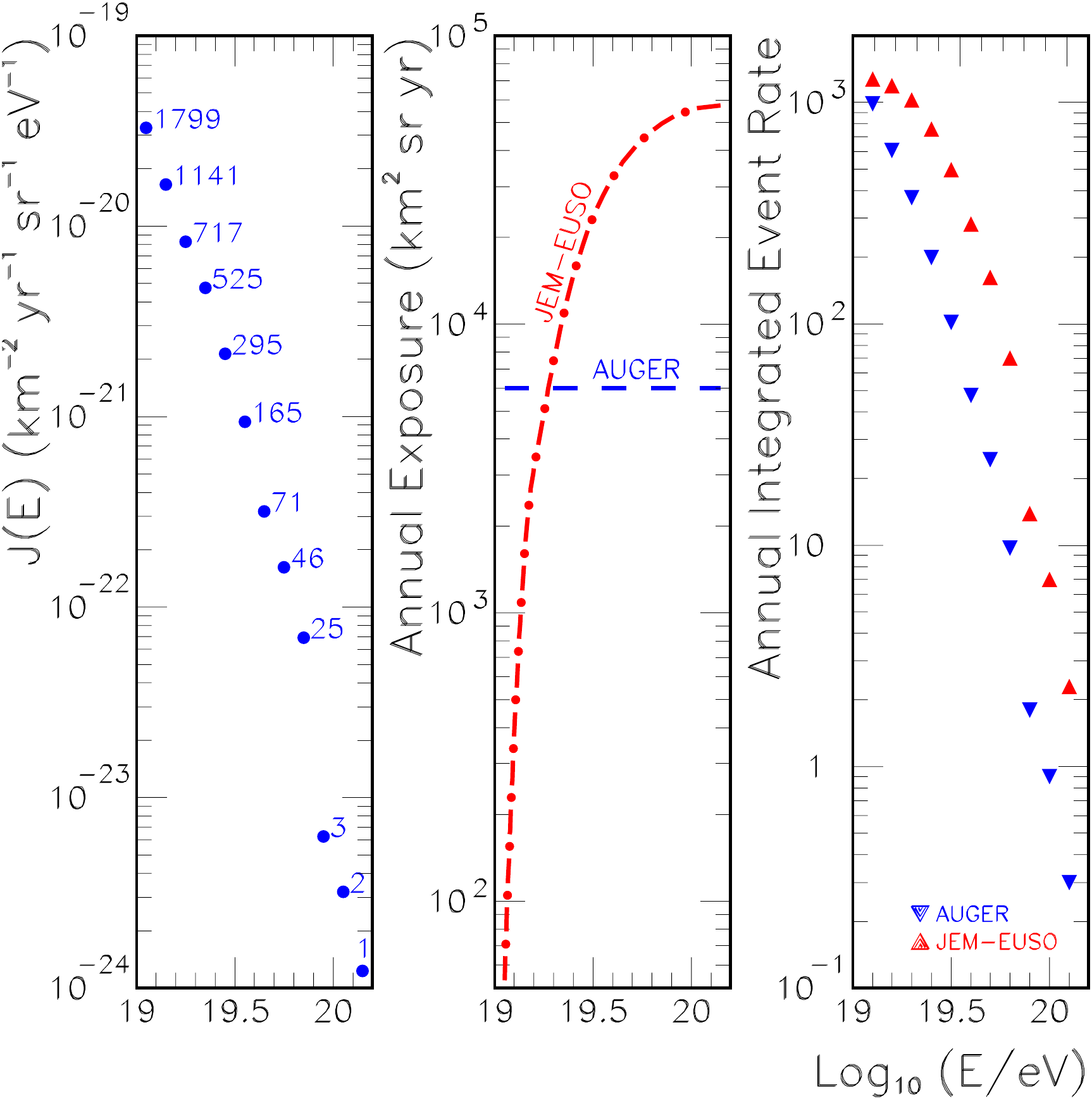}
\caption{Left panel: The most recent energy spectrum reported by the
  Pierre Auger Collaboration~\cite{Abreu:2011pj}. Middle panel: Annual
  exposure of JEM-EUSO~\cite{Shinozaki} and
  Auger~\cite{Abraham:2010zz}. Right panel: Estimated event rates
  derived on the basis of the spectrum in the left panel.}
\label{exposure-rates}
\end{center}
\end{figure}

\begin{figure}[t]
\begin{center}
\includegraphics[width=0.48\textwidth]{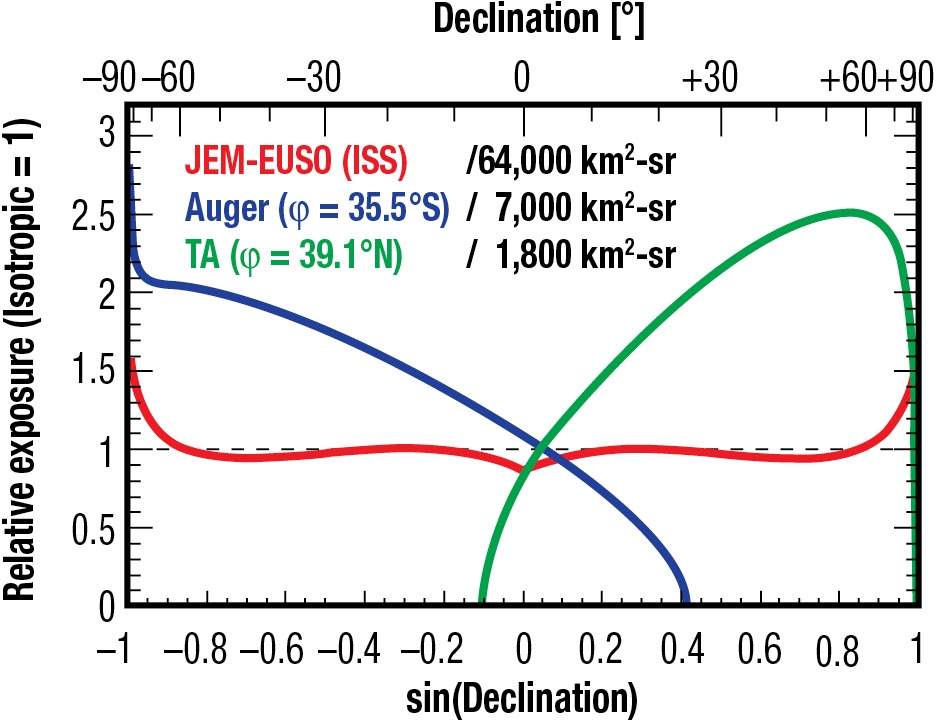}
\caption{Normalized relative exposure, $\omega(\sin \delta)$, of
  Auger, TA, and JEM-EUSO. The latter has a nearly
  uniform exposure over most of the sky with only small excesses in
  polar regions.}
\label{RE}
\end{center}
\end{figure}

\begin{figure}[t]
\begin{center}
\includegraphics[width=0.47\textwidth]{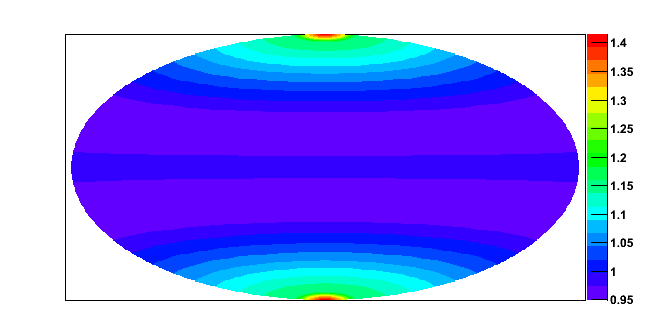}
\caption{Sky map in equatorial coordinates showing the relative exposure of JEM-EUSO.} 
\label{RE-euso}
\end{center}
\end{figure}

The observation area of JEM-EUSO depends upon the tilting angle $\xi$
off the nadir and $h_{\rm ISS}$. For the baseline layout, the
observation area for nadir mode is given by
\begin{equation}
A_{\rm obs}^{({\rm nadir})} \approx 1.4 \times 10^5 \left(\frac{h_{\rm ISS}}{400~{\rm km}}\right)~{\rm km}^2 \, .
\end{equation} 
For tilting angles  $\xi \alt 40^\circ$, 
\begin{equation}
A_{\rm obs} (\xi) \approx A_{\rm obs}^{({\rm nadir})} \, (\cos \xi)^{-b} \,,
\end{equation}
where, for the altitude of interest,   $3.2 \leq b \leq 3.4$.  For $\xi
\sim 40^\circ - 50^\circ$, a part of the FOV views the sky over the local
horizon, and $A_{\rm obs}$ saturates above $\xi \sim 60^\circ$.

The UV tracks of EAS must be discriminated in the nightglow
background. One key parameter is therefore the fraction of time in
which EAS observations are not hampered by the brightness of the
sky. The observational duty-cycle $\eta_0$ is defined as the fraction
of time in which the sky is dark enough to measure EAS. By rescaling
measurements of the Tatiana satellite to the ISS orbit, $\eta_0$ is
estimated to be 19\%~\cite{Bobik,Adams:2012tt}. In this estimate all major
atmospheric effects, such as lightnings, meteors, and anthropic lights
(e.g. city lights) have been included.

The impact of clouds is estimated by the global secular statistics of
the optical depth and cloud-top altitude~\cite{Garino,Adams:2012tt} convolved with the
trigger probability for each case.  Applying quality cut for events
with shower maximum above the optically thick clouds, this efficiency
 factor for $E > 3 \times 10^{19}~{\rm eV}$ is estimated to be 
$\kappa_{\rm c} \sim 70\%$ (see~\cite{SaezCano,Adams:2012tt} for details).

The first step to computing the JEM-EUSO geometric aperture involves
an estimate of the trigger efficiency.  This calculation is performed
using a dedicated detector simulation package~\cite{Berat:2009va,Fenu,Adams:2012tt},
together with a fast shower simulation allowing for large statistics.
Estimates assume the fluorescence yield reported 
in~\cite{Nagano:2004am}.  The
geometric aperture ${\cal A}$ is then (conservatively) computed for
nadir mode by folding the trigger efficiency with geometrical
acceptance for showers of various energies and incident angles, 
assuming a clear sky condition with average background
level~\cite{Shinozaki,Adams:2012tt}. The exposure growth per unit time is given by
\begin{equation}
\frac{d{\cal E}}{dt} = {\cal A} \, \, \eta_0 \,\, \kappa_{\rm c} \, .
\end{equation}
In Fig.~\ref{exposure-rates} we show the annual exposure (growth in
exposure by one year of operation) together with expected event rate
on the basis of the energy spectrum reported by the Pierre Auger
Collaboration~\cite{Abreu:2011pj}. This {\em conservative} event rate
will be taken as fiducial throughout this report. However, it is
important to keep in mind that the expected event rate would increase
by about a factor of 2 (see Fig.~\ref{fig:cal}) if the spectrum
reported by the HiRes Collaboration~\cite{Abbasi:2007sv} is taken for
normalization of the CR intensity.

The scientific requirements of the mission~\cite{Santangelo:2010zz}
can be summarized as follows:
\begin{itemize}
\item statistical uncertainty on the energy measurement: $\Delta E/E <
  30\%$ at $E = 10^{20}~{\rm eV}$ and $60^\circ$ zenith angle;
\item observation area: $A_{\rm obs}^{({\rm nadir})} \geq 1.3 \times 10^5~{\rm km^2}$   for an orbit height $h_{\rm ISS} \approx 400~{\rm km}$;
\item arrival direction accuracy:
better than $2.5^\circ$ for $60^\circ$ zenith angle and  
$E = 10^{20}~{\rm eV}$;
\item accuracy on $X_{\rm max}$: $\Delta X_{\rm max}
  \sim 120~{\rm g} \, {\rm cm}^{-2}$ at primary energy $E= 10^{20}~{\rm eV}.$
\end{itemize}
Two independent  end-to-end simulation packages,
``Saitama'' developed in Japan and the EUSO Simulation and Analysis
Framework (ESAF)~\cite{Berat:2009va} developed in Europe, have been used to evaluate the scientific performances of the instrument. In all cases the
simulations show that JEM-EUSO will be able to meet these
science requirements~\cite{Santangelo:2011zz}.

  Since the ISS orbits the Earth in the latitude range $\pm 51^\circ$,
  moving at a speed of $\sim 7~{\rm km/s}$, JEM-EUSO will monitor both
  hemispheres with a rather uniform exposure (see Figs.~\ref{RE} and
  \ref{RE-euso}). This will reduce systematic uncertainties in 
anisotropy studies. We discuss this next.

\section{Anisotropy in Arrival Directions}
\label{sec:ani}

The detection of anisotropies in the distribution of arrival directions will  
yield the strongest clues to the  origin of UHECRs.  If cosmic rays are observed to
cluster within a small angular region or show
directional alignment with powerful compact objects, one might be able
to associate them with isolated sources in the sky. Alternatively, 
if the distribution of arrival directions exhibits a large-scale
anisotropy, this would indicate certain classes of
sources that are associated with observed large-scale structure (such as the
Galactic plane, the Galactic halo, nearby galaxy groups, or the local Supercluster of galaxies).  

The lack of clustering on small scales of the currently known
118 events around the extreme energy range
(98  from Auger with $E > 5.5 \times
10^{19}$~eV~\cite{Olinto:2012dc} while the Telescope
Array reports 20 events with $E > 5.7 \times 10^{19}$~eV~\cite{Ikeda})
may indicate that either sources are transient, have large number
densities, and/or magnetic fields still deflect these particles
significantly at these energies~\cite{Abreu:2011md}. On larger scales, there are hints of
anisotropies in the observations of the Pierre Auger Observatory. With
the 98  events accumulated until June 2011, Auger
reports a 3$\sigma$ departure from isotropy using a prescription that
compares arrival directions with the distribution of nearby active
galactic nuclei~\cite{Olinto:2012dc}. The Telescope Array results are
consistent with this level of anisotropy. In particular, the most
prominent feature in the UHECR sky is an excess of events
in the vicinity of the nearby active galaxy Centaurus A (Cen A).

Anisotropy studies strongly depend on the
primary particle species. The smaller magnetic rigidity of heavy
nuclei can relax a critical problem faced by current observations, the
lack of a suitable clear source in the sky. The lower rigidity
(smaller Larmor radius) can lead to significant deflections in typical
extragalactic magnetic fields, ${\cal O} ({\rm nG})$, and would
postpone the expected anisotopies in the arrival directions to larger
energies.\footnote{Although not generally appreciated, it is important
  to note that as a statistical average over the sky, an all pervading
  extragalactic magnetic field is constrained to be $B \alt 3 \times
  10^{-7} (\Omega_b h^2/0.02)^{-1} \, (h/0.72) \, (\lambda/{\rm
    Mpc})^{1/2}~{\rm G}$, where $\lambda$ is the coherent length,
  $\Omega_b h^2 \simeq 0.02$ is the baryon density, and $h \simeq
  0.72$ is the present day normalized Hubble expansion
  rate~\cite{Blasi:1999hu,Farrar:1999bw}. This upper limit, derived
  from measurements of the Faraday rotation in the linearly polarized
  radio emission from distant quasars~\cite{Kronberg:1993vk}, depends
  significantly on assumptions about the electron density profile as a
  function of redshift~\cite{Blasi:1999hu,Farrar:1999bw}.  Because
  $\Omega_b$ has contributions from neutrons and only electrons in
  ionized gas are relevant to Faraday rotation, the previous estimate
  should be taken as a conservative bound. If 
  extragalactic magnetic fields saturate this upper limit, even EECR
  protons may suffer large deflections {\em en route} to
  Earth~\cite{Farrar:1999bw}.}

Ground-based air shower detectors which experience stable operation
over a period of a year or more will have an uniform exposure in right
ascension, $\alpha$. In such a case, the right ascension distribution
of the UHECR flux arriving at a detector can be characterized by the
amplitudes and phases of its Fourier expansion,
\begin{equation}
I (\alpha) = I_0 [1 + r \cos (\alpha - \phi)  + r' \cos (2 (\alpha - \phi')) + \dots \,] \, .
\end{equation}
For $N$ measurements $\alpha_i$, the first harmonic amplitude $r$ and
its phase $\phi$ can be determined by applying the classical
Rayleigh formalism~\cite{Linsley:1975kp},
\begin{equation}
r = \sqrt{x^2 + y^2}\,, \quad \quad \quad \quad \phi = {\rm arctan} \frac{y}{x} \, ,
\end{equation}
where
\begin{equation}
x = \frac{2}{{\cal N}} \sum_{i=1}^{N}  \, w_i \, \cos \alpha_i \,, \,\,\,\,\,y =
\frac{2}{{\cal N}} \sum_{i=1}^{N}  \, w_i\,\,
\, \sin \alpha_i\,,
\label{eqn:fh}
\end{equation}
${\cal N} = \sum_{i=1}^{N} w_i$ is the normalization factor, and the
weights, $w_i = \omega^{-1}(\delta_i)$, are the reciprocal of the
relative exposure, $\omega$, given in Fig.~\ref{RE} as a function of the declination,
$\delta_i$. As deviations from an uniform right
ascension exposure are small, the probability $P(>r)$ that an
amplitude equal or larger than $r$ arises from an isotropic 
distribution can be approximated by the cumulative distribution
function of the Rayleigh distribution $P(>r) = \exp (-k_0),$ where $k_0
= {\cal N} \, r^2/4$. 

The first harmonic amplitude of the  right ascension distribution
can be directly related to the amplitude $\alpha_d$ of a dipolar
distribution of the form 
\begin{equation}
J(\alpha,\delta) = (1+ \alpha_d \ \hat d
\cdot \hat u) \,  J_0 \,, 
\end{equation}
where $\hat u$ denotes the unit vector in the opposite direction of the shower arrival direction
and $\hat d$  denotes the
unit vector in the direction of the dipole. We can rewrite $x$, $y$, and
$\mathcal{N}$ as
\begin{eqnarray}
x&=& \frac{2}{\mathcal{N}} \int_{\delta_{\rm min}}^{\delta_{\rm max}} d\delta \int_0^{2\pi}
d\alpha \cos \delta \ J(\alpha,\delta) \ \omega(\delta) \cos \alpha, \nonumber \\  
y&=& \frac{2}{\mathcal{N}} \int_{\delta_{\rm min}}^{\delta_{\rm max}} d\delta \int_0^{2\pi}
d\alpha \cos \delta \ J(\alpha,\delta) \ \omega(\delta) \sin \alpha,  \label{corolo} \\ 
\mathcal{N}&=& \int_{\delta_{\rm min}}^{\delta_{\rm max}} d\delta \int_0^{2\pi}
d\alpha \cos \delta \ J(\alpha,\delta) \ \omega(\delta) \, .\nonumber
\end{eqnarray}
In (\ref{corolo}) we have neglected  the small dependence on right 
ascension in the exposure. Next, we write the angular dependence in 
$J(\alpha,\delta)$ as 
\begin{equation}
 \hat d \cdot \hat u_i = \cos \delta_i \cos \delta_0 
\cos (\alpha_i-\alpha_0) + \sin \delta_i \sin \delta_0 \,,
\end{equation}
where $\alpha_0$ and $\delta_0$ are the right ascension and
declination of the apparent origin of the dipole, and $\alpha_i$ and $\delta_i$ are the right ascension and declination of the $i$th event. Performing the $\alpha$ integration in (\ref{corolo}) it
follows that
\begin{equation} 
\label{eqn:amplitudes}
r=\left| \frac{A \alpha_d^\perp}{1+B \alpha_d^\parallel} \right|
\end{equation}
where $\alpha_d^\parallel=\alpha_d\sin{\delta_0}$ is the component of the dipole
along the Earth rotation axis, and $\alpha_d^\perp=\alpha_d
\cos{\delta_0}$ is the component in the equatorial
plane~\cite{Aublin:2005am}. The coefficients $A$ and $B$ can be
estimated from the data as the mean values of the cosine and the sine
of the event declinations,
$$
A =\frac{\int d\delta\,\omega(\delta) \cos^2 \delta}
{\int d\delta\,\omega(\delta) \cos \delta} \,, \quad \quad B
=\frac{\int d\delta\,\omega(\delta) \cos \delta \sin \delta} {\int
  d\delta\,\omega(\delta) \cos \delta} \,;
$$
e.g., for the Auger data sample $A = \langle \cos \delta \rangle \simeq 0.78$ and $B = \langle \sin \delta \rangle \simeq -0.45$~\cite{Abreu:2011zzj}.
For a dipole amplitude $\alpha_d$, the
measured amplitude of the first harmonic in right ascension $r$ thus
depends on the region of the sky observed, which is essentially a
function of the latitude of the observatory $\varphi$, and the
range of zenith angles considered. In the case of a small $B
\alpha_d^\parallel$ factor, the dipole component in the equatorial
plane is obtained as $\alpha_d^\perp\simeq r/A$.  The
phase $\phi$ corresponds to the right ascension of the dipole
direction $\alpha_0$.\footnote{A point worth noting at this juncture: A pure dipole distribution is
not possible because the cosmic ray intensity cannot be negative in
half of the sky.  A ``pure dipole deviation from isotropy'' means a
superposition of monopole and dipole, with the intensity everywhere
$\geq 0$. An approximate dipole deviation from isotropy could be
caused by a single strong source if magnetic diffusion or dispersion
distribute the arrival directions over much of the sky. However, a
single source would produce higher-order moments as well.}

Recently, an intriguing concentration of events has been reported by
the Pierre Auger Collaboration~\cite{:2010zzj} in the region around
the direction of Cen A, a powerful radiogalaxy at a distance of
3.4~Mpc with equatorial coordinates $(\alpha_0,\, \delta_0) =
(201.4^\circ, -43.0^\circ)$.\footnote{The potential of Cen A to
  accelerate both protons and heavy nuclei in the outer regions of the
  giant radio lobes has been recently discussed
  in~\cite{Peer:2011xf}.} Out of the 69 Auger events with $E >
5.5 \times 10^{19}~{\rm eV}$ (collected over 6~yr but equivalent to
2.9~yr of the nominal exposure/yr of the full Auger), the maximum
departure from isotropy occurs for a ring of $18^\circ$ around the
object, in which 13 events are observed compared to an expectation of
3.2 from isotropy. The anisotropy amplitude obtained from the 69
arrival directions, assuming a dipole function for a source model with
a maximum value at Cen A is found to be $\alpha_d \sim
0.25$~\cite{Anchordoqui:2011ks}.\footnote{If the observed excess is
  produced by heavy nuclei accelerated at Cen A, then a larger
  anisotropy would be expected at $E \sim 10^{18}~{\rm
    eV}$~\cite{Lemoine:2009pw} . The lack of ``low energy'' anisotropy
  in the Auger data sample constrains the spectral index of the
  potential nucleus emission spectrum~\cite{Abreu:2011vm}.} It is
important to keep in mind that this is an {\em a posteriori} study, so
{\em one cannot use it to determine a confidence level for anisotropy
  as the number of trials is unknown}.

The right ascension harmonic analyses are completely blind to
intensity variations which depend only on declination.  Combining
anisotropy searches in right ascension over a range of declinations
could dilute the results, since significant but out of phase Rayleigh
vectors from different declination bands can cancel each other out.  A
cosmic ray detector with full-sky coverage like JEM-EUSO can exploit
standard anisotropy analysis methods that do not work if part of the
celestial sphere is never seen. In particular, the distribution of
arrival directions can be fully characterized by a set of spherical
harmonics coefficients~\cite{Sommers:2000us,Anchordoqui:2003bx}.
These coefficients for a function on a sphere are the analogue of
Fourier coefficients for a function on a plane. Variations on an
angular scale of $\theta$ radians contribute amplitude in the $\ell =
1/\theta$ modes just as variations of a plane function on a distance
scale of $\lambda$ contribute amplitude to the Fourier coefficients
with $k=2\pi/\lambda$. 

For anisotropy searches at JEM-EUSO, we might look for power in modes
from $\ell = 1$ (dipole) out to $\ell \sim 20$, higher order modes
being irrelevant because the detector will smear out any true
variations on scales that are smaller than its angular resolution. The
dipole can be recovered from the celestial intensity function by
\begin{equation}
\alpha_d \, \hat{d} = \frac{3}{\cal N}\\ \int J(\hat{u})\ \hat{u}\
d\Omega \, .
\end{equation}
The components of the dipole vector
are found to be
\begin{equation}
\alpha_d \, d_a = \frac{3}{\cal N}\ \sum_{i=1}^{N}\
\frac{1}{\omega_i}\ u^{(i)}_a, 
\end{equation}
where $u^{(i)}_a$ denotes a component of the $i$th vector.  (These
dipole components are linear combinations of the three spherical
harmonic coefficients with $\ell=1$.)

The sensitivity of JEM-EUSO to a dipole of amplitude $\alpha_d$ can 
then be estimated via simulation of an
ensemble of artificial data sets  (with random dipole
directions $\hat{d}$).  For each data set, we use the above formula to
determine the dipole vector, and record the difference of the estimated
$\alpha_d$ from the input dipole amplitude and also the angle between the
estimated direction and the input dipole direction.  These error
distributions describe the measurement accuracy.  The RMS deviation
from the true $\alpha_d$ is a single number to characterize the
amplitude measurement accuracy, and the average space angle error
summarizes the accuracy of determining the dipole direction.  For a
fixed number of arrival directions $N$, the RMS error in the amplitude has
little dependence on the amplitude. For the purpose of detecting an
anisotropy (as opposed to measuring it), the relevant quantity is the
amplitude divided by the RMS error, which is the number of sigmas
deviation from isotropy~\cite{Sommers:2000us}
\begin{equation}
\frac{\alpha_d}{\Delta \alpha_d} \approx 0.65 \, \alpha_d \, \sqrt{N} \, .
\end{equation}
It is straightforward to see that {\em after 3~yr of operation the
  JEM-EUSO mission will collect about 410 events above $5.5 \times
  10^{19}$~eV providing an ``a priori'' test for a dipole anisotropy
  with amplitude $\alpha_d = 0.25$ (for a source model with a maximum
  value at Cen A) at the $3.3\sigma$ level. If the observed excess of
  events recorded by the Pierre Auger Observatory is not a statistical
  fluctuation, in 5~yr of JEM-EUSO  operations the significance would
  be increased up to $4.2\sigma$}. To reach the standard $5\sigma$
discovery level about $1,000$ events would be required.

\section{Neutrino Sensitivity}

In addition to studying the highest energy cosmic rays, JEM-EUSO is
also capable of observing extreme energy cosmic neutrinos
(EEC$\nu$'s). At present, the IceCube telescope at the South Pole
holds the record for the most energetic neutrino interactions
observed~\cite{Abbasi:2010ie}; these events have energies up to $\sim
10^{14}$~eV and are consistent with the predicted spectrum of
atmospheric neutrinos. JEM-EUSO, by contrast,  can  detect
neutrinos with energies $E_\nu \agt 10^{20}$~eV.

The flux of high energy ($E_\nu > 10^{12}~{\rm eV}$) neutrinos at
Earth is expected to be very faint and their interactions with matter
are very rare (the interaction length of multi-TeV neutrinos is
of the order of the Earth's diameter). Therefore, neutrino telescopes
have to face the enormous challenge of observing and identifying very
rare neutrino interactions in huge detection volumes~\cite{Anchordoqui:2009nf}. Secondary
charged particles produced in weak interactions of neutrinos with
nuclei can be identified by Cherenkov light emission in optically
transparent media. This method has been successfully applied in Lake
Baikal~\cite{Aynutdinov:2005dq} , the Mediterranean
(ANTARES~\cite{:2010ec}) and the Antarctic glacier
(AMANDA~\cite{Achterberg:2007qp,Ackermann:2007km},
IceCube~\cite{Abbasi:2011ji,Abbasi:2011jx}). Coherent radio Cherenkov
emission has been studied from the regolith of the Moon
(GLUE~\cite{Gorham:2003da}), in the Greenland ice sheet
(FORTE~\cite{Lehtinen:2003xv}) and in the Antarctic ice
(RICE~\cite{Kravchenko:2006qc},
ANITA~\cite{Gorham:2008yk,Gorham:2010kv}).

At sufficiently high energies cosmic neutrinos can trigger atmospheric
air showers similar to those due to high energy cosmic rays (baryons
or photons). Although greatly reduced by $\sigma_{\nu N} \approx 6
\times (E_\nu/{\rm GeV})^{0.358}~{\rm pb}$ at $E_\nu \agt 10^{20}~{\rm
  eV}$~\cite{Anchordoqui:2006ta}, the neutrino interaction length
\begin{equation}
  L = 1.7 \times 10^7 \, \left(\frac{\rm pb}{\sigma_{\nu N}} \right)~{\rm kmwe}
\end{equation}
is far larger than the Earth's atmospheric depth, which is only
0.36~kmwe even when traversed horizontally. Neutrinos therefore
produced EAS uniformly at all atmospheric depths. As a result, the
most promising neutrino signal are quasi-horizontal showers initiated
deep in the atmosphere~\cite{Capelle:1998zz}. For showers with large enough zenith angles,
the likelihood of interaction is maximized and the background from
baryonic cosmic rays is eliminated, since these  showers begin high in the
atmosphere. Ground-based cosmic ray observatories
(HiRes~\cite{Abbasi:2008hr},
Auger~\cite{Abraham:2007rj,Abraham:2009uy,Abreu:2011zz}, and TA) can also
identify neutrino events as electromagnetic showers from
Earth-skimming
tau-neutrinos~\cite{Feng:2001ue,Bertou:2001vm,Fargion:2000iz}. As
 explained in Sec.~\ref{sec:jem-euso}, neutrino
induced showers recorded by the JEM-EUSO mission can also be
distinguished from those of protons and nuclei by requiring that they
be {\em deeply penetrating}~\cite{Supanitsky:2011jp}.

\begin{figure}[htbp]
\begin{center}
\includegraphics[width=0.47\textwidth]{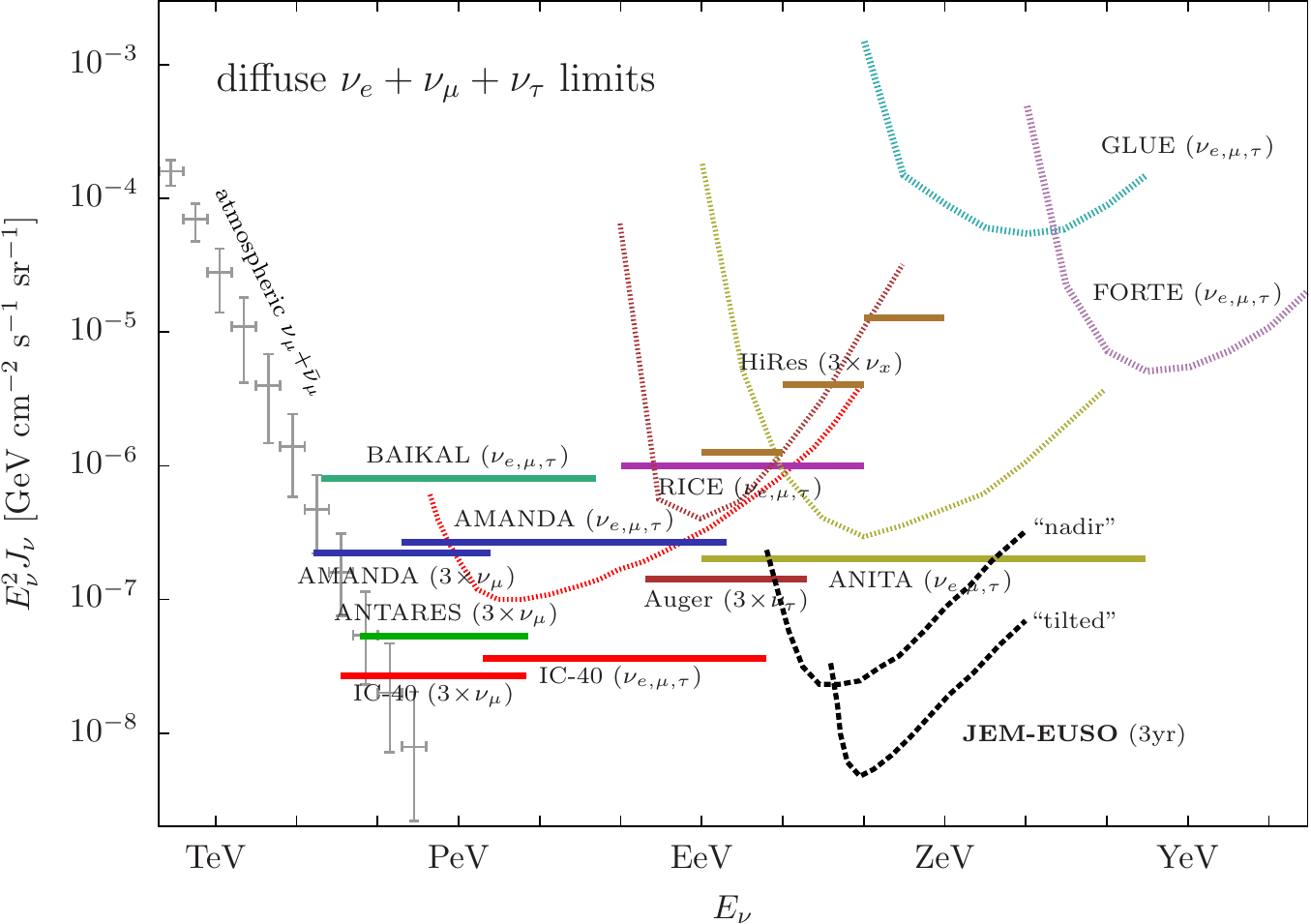}
\caption{Summary of diffuse neutrino
  limits~\cite{Aynutdinov:2005dq,:2010ec,Achterberg:2007qp,Ackermann:2007km,Abbasi:2011ji,Abbasi:2011jx,Gorham:2003da,Lehtinen:2003xv,Kravchenko:2006qc,Gorham:2008yk,Gorham:2010kv,Abbasi:2008hr,Abraham:2007rj,Abraham:2009uy,Abreu:2011zz}
  and the flux of atmospheric neutrinos~\cite{Abbasi:2010ie}. For
  comparison, the EEC$\nu$ flux sensitivity of JEM-EUSO (for nadir and tilted
  modes) to detect 1 event/energy-decade/3~yr
  is also shown~\cite{Takahashi:2009zzc}.}
\label{nus}
\end{center}
\end{figure}

Despite the large experimental effort, neutrino telescopes have yet to
identify the first extragalactic neutrino source and so far only place
upper bounds on their fluxes. In Fig.~\ref{nus} we show a summary of
diffuse neutrino limits from the various experiments mentioned
above. In combining these limits we assume that the total neutrino
flux arrives at Earth with an equal composition of flavors. This is
expected from neutrino production via pion decay in weak magnetic
fields and subsequent flavor oscillations over cosmological distances.

In addition  to the direct neutrino limits shown in Fig.~\ref{nus},
there are various indirect limits arising from the so-called
\mbox{``CR $\leftrightharpoons \nu$} connection.''  We discuss these
next.

It is helpful to envision the CR engines as machines where
protons are accelerated and (possibly) permanently confined by the
magnetic fields of the acceleration region. The production of neutrons
and pions and subsequent decay produces neutrinos, $\gamma$-rays, and
CRs. If the neutrino-emitting source also produces high energy
CRs, then pion production must be the principal agent for the
high energy cutoff on the proton spectrum.  Conversely, since the
protons must undergo sufficient acceleration, inelastic pion
production needs to be small below the cutoff energy; consequently,
the plasma must be optically thin. Since the interaction time for
protons is greatly increased over that of neutrons due to magnetic
confinement, the neutrons escape before interacting, and on decay give
rise to the observed CR flux. The foregoing can be summarized
as three conditions on the characteristic nucleon interaction time
scale $\tau_{\rm int}$; the neutron decay lifetime $\tau_n$; the
characteristic cycle time of confinement $\tau_{\rm cycle}$; and the
total proton confinement time $\tau_{\rm conf}$: $(i)\; \tau_{\rm
  int}\gg \tau_{\rm cycle}$; $(ii)\; \tau_n > \tau_{\rm cycle}$;
$(iii)\; \tau_{\rm int}\ll \tau_{\rm conf}$. The first condition
ensures that the protons attain sufficient energy.  Conditions $(i)$
and $(ii)$ allow the neutrons to escape the source before
decaying. Condition $(iii)$ permits sufficient interaction to produce
neutrons and neutrinos. These three conditions together define an
optically thin source~\cite{Ahlers:2005sn}. A desirable property to
reproduce the almost structureless energy spectrum is that a single
type of source will produce cosmic rays with a smooth spectrum across
a wide range of energy.

The cosmic ray flux above the ankle is often
summarized as ``one $3 \times 10^{10}$~GeV particle per kilometer square per
year per steradian.'' This can be translated into an energy flux~\cite{Gaisser:1997aw}
\begin{eqnarray}
E \left\{ E \, J (E)\right\} & = & {3 \times 10^{10}\,{\rm GeV} 
\over \rm (10^{10}\,cm^2)(3\times 10^7\,s) \, sr} \nonumber \\
 & = &  10^{-7}\rm\, GeV\ cm^{-2} \, s^{-1} \, sr^{-1} \,.
\end{eqnarray}
From this we can derive the energy density $\epsilon_{\rm CR}$ in UHECRs using flux${}={}$velocity${}\times{}$density, or
\begin{equation}
4\pi \int  dE \left\{ E \, J(E) \right\} =  c \, \epsilon_{\rm CR}\,.
\end{equation}
This leads to
\begin{equation}
  \epsilon_{\rm CR} = {4\pi\over c} \int_{E_{\rm min}}^{E_{\rm max}} { 10^{-7}\over E} 
  dE \, {\rm {GeV\over cm^2 \, s}} \simeq   10^{-19} \, 
{\rm {TeV\over cm^3}} \,,
\end{equation}
taking the extreme energies of the accelerator(s) to be $E_{\rm min} \simeq 10^{19}~{\rm eV}$ and
$E_{\rm max}  = 10^{21}~{\rm eV}$. The power required for a population of  sources
to generate this energy density over the Hubble time (${\cal T}_H \approx 10^{10}$~yr) is: 
$\dot \epsilon_{\rm CR}^{[10^{10}, 10^{12}]} \sim 5 \times 10^{44}~{\rm TeV} \, {\rm Mpc}^{-3} \, {\rm yr}^{-1} \simeq 3 \times 10^{37}~{\rm erg} \, {\rm Mpc}^{-3} \, {\rm s}^{-1}$.  This works out to roughly
({\it i}\,) $L \approx 3 \times 10^{39}$ erg s${}^{-1}$ per galaxy, 
({\it ii}\,) $L \approx 3 \times 10^{42}$ erg s${}^{-1}$ per cluster of galaxies, 
({\it iii}\,) $L \approx 2 \times 10^{44}$ erg s${}^{-1}$ per active galaxy, or 
({\it iv}\,) $\int L \, dt\approx  10^{52}$ erg per cosmological
gamma-ray burst~\cite{Gaisser:1997aw}. 
The coincidence between these numbers and
the observed output in electromagnetic energy of these sources
explains why they have emerged as the leading candidates for the
CR accelerators. 

The energy production rate of protons derived professionally, assuming a cosmological distribution of  sources (with injection spectrum  typical of shock acceleration $dN_0/dE \propto E^{-2}$) is~\cite{Waxman:1995dg} 
\begin{equation}
\dot \epsilon_{\rm CR}^{[10^{10}, 10^{12}]} \sim 5 \times 10^{44}~{\rm erg} \, {\rm Mpc}^{-3} \, {\rm yr}^{-1} \, .
\label{professionally}
\end{equation}
This is within a factor of a few of  our back-of-the-envelope estimate (1 TeV = 1.6 erg). The energy-dependent generation rate of CRs is therefore given by 
\begin{eqnarray}
E^2  \frac{d \dot n}{dE } 
 & = & \frac{\dot \epsilon_{\rm CR}^{[10^{10}, 10^{12}]}}{\ln(10^{12}/10^{10})} \nonumber \\
& \approx & 10^{44}\,\rm{erg}\,\rm{Mpc}^{-3} \rm{yr}^{-1} \,\, .
\end{eqnarray} 

The energy density of neutrinos produced through $p\gamma$
interactions of these protons can be directly tied to the injection
rate of CRs
\begin{equation}
E^2_{\nu} \frac{dn_{\nu}}{dE_{\nu}}
\approx \frac{3}{8} \epsilon_\pi \, {\cal T}_H \,E^2 \, \frac{d \dot n}{dE} \,,
\end{equation}
where  $\epsilon_\pi$ is the
fraction of the energy which is injected in protons lost into photopion
interactions.  The factor of 3/8 comes from the fact that, close to
threshold, roughly half the pions produced are neutral, thus not
generating neutrinos, and one quarter of the energy of charged pion
decays goes to electrons rather than neutrinos. Namely, resonant $p \gamma$ interactions produce twice as many neutral pions as charged pions. Direct pion production via virtual  meson exchange contributes only about 20\% to the total cross section, but is almost exclusively into $\pi^+$. Hence, $p \gamma$ interactions produce roughly equal numbers of $\pi^+$ and 
$\pi^0$. The average  neutrino energy from the direct pion decay is $\langle E_{\nu_\mu}
  \rangle^\pi = (1-r)\,E_\pi/2 \simeq 0.22\,E_\pi$ and that of the
  muon is $\langle E_{\mu} \rangle^\pi = (1+r)\,E_\pi/2 \simeq
  0.78\,E_\pi$, where $r$ is the ratio of muon to the pion mass
  squared. In muon decay, since the $\nu_\mu$ has about 1/3 of its parent energy,
  the average muon neutrino energy is $\langle E_{\nu_\mu} \rangle^\mu =(1+r)E_\pi/6=0.26 \,
  E_\pi$.

The ``Waxman-Bahcall (WB) bound'' is defined by the condition $\epsilon_\pi
=1$
\begin{eqnarray} 
E^2_{\nu} \ J^{\rm WB}_{{\rm all} \, \nu}  & \approx & (3/8)
  \,\xi_z\, \epsilon_\pi\, {\cal  T}_H \, \frac{c}{4\pi}\,E^2 \, 
   \frac{d \dot n}{dE}  \nonumber \\ & \approx & 2.3
  \times 10^{-8}\,\epsilon_\pi\,\xi_z\, \rm{GeV}\,
  \rm{cm}^{-2}\,\rm{s}^{-1}\,\rm{sr}^{-1},
\label{wbproton}
\end{eqnarray}
where the parameter $\xi_z$ accounts for the effects of source
evolution with redshift, and is expected to be $\sim
3$~\cite{Waxman:1998yy}. For interactions with the ambient gas ({\em i.e.} 
$pp$ rather than $p \gamma$ collisions), the average fraction of the
total pion energy carried by charged pions is about $2/3$, compared to
$1/2$ in the photopion channel. In this case, the upper bound given
in Eq.~(\ref{wbproton}) is enhanced by 33\%.  

The actual value of the neutrino flux depends on what fraction of the
proton energy is converted to charged pions (which then decay to
neutrinos), {\em i.e.}  $\epsilon_\pi$ is the ratio of charged pion energy to the {\em
  emerging} nucleon energy at the source.  For resonant
photoproduction, the inelasticity is kinematically determined by
requiring equal boosts for the decay products of the
$\Delta^+$, giving $\epsilon_\pi = E_{\pi^+}/E_n
\approx 0.28$, where $E_{\pi^+}$ and $E_n$ are the emerging charged pion
and neutron energies, respectively.  For $pp\rightarrow NN + {\rm
  pions},$ where $N$ indicates a final state nucleon, the inelasticity
is $\approx 0.6$~\cite{Alvarez-Muniz:2002ne}. This then implies that
the energy carried away by charged pions is about equal to the
emerging nucleon energy, yielding (with our definition)
$\epsilon_\pi\approx 1.$

At production, if all muons decay, the neutrino flux consists of equal
fractions of $\nu_e$, $\nu_{\mu}$ and $\bar{\nu}_{\mu}$. Originally,
the WB bound was presented for the sum of $\nu_{\mu}$ and
$\bar{\nu}_{\mu}$ (neglecting $\nu_e$), motivated by the fact that
only muon neutrinos are detectable as track events in neutrino
telescopes. Since oscillations in the neutrino sector mix the
different species, we chose instead to discuss the sum of all neutrino
flavors. When the effects of oscillations are accounted for, {\it
  nearly} equal numbers of the three neutrino flavors are expected at
Earth~\cite{Learned:1994wg}.

$\gamma$-ray telescopes also constrain the diffuse flux of neutrinos.
High energy $\gamma$-rays produced in distant sources cannot reach
Earth unscathed. At multi-TeV energies $\gamma$-rays interact with
universal radiation fields, producing $e^+ e^-$ pairs. The
$\gamma$-radiation is recycled to somewhat lower energies by inverse
Compton scattering of the electrons and positrons off the same
radiation background. These two mechanisms develop electromagnetic
cascades until the center-of-mass energy of $\gamma \gamma$-scattering
drops below the pair production (PP) threshold at the order of MeV. For
optical photons with energies of the order of eV this occurs at TeV
energies. Other processes like synchrotron radiation in extragalactic
magnetic fields, double or triple pair production can also contribute
to the cascades spectrum. The net result is a pile up of $\gamma$-rays
at GeV-TeV energies.

The {\em cascade limit}  of diffuse neutrino fluxes is a consequence
of the bolometric energy budget of this process~\cite{Berezinsky:1975zz}. The inferred energy density $\omega_\gamma$ of the extragalactic diffuse $\gamma$-ray background in the GeV-TeV region constitutes an upper limit for the total electromagnetic energy from pion production of CR protons. 

If the energy loss of pions prior to decay is neglible the neutrino energy relates to the $\gamma$-ray energy as $E_\nu \simeq E_\gamma/2$ (per neutrino). The total $\nu$ and $\gamma$ emissivity ($Q$ in units of ${\rm GeV}^{-1} {\rm  s}^{-1}$) from pion decays can then be related by particle number conservation as
\begin{equation}
Q_{{\rm all} \, \nu} (E_\nu) = \frac{3}{2} \, \frac{\Delta E_\gamma}{\Delta E_\nu} \, K Q_\gamma(E_\gamma) \,,
\label{Q}
\end{equation}
where the factor $K$ depends on the relative multiplicities of charged
and neutral pions, $K = N_{\pi^\pm}/N_{\pi^0}.$ There is also one
electron from the decay of the muon which contributes as $Q_{{\rm all}
  \, \nu} \simeq 3 Q_e(E_\nu)$. From Eq.~(\ref{Q}) with $K=1$ we have
\begin{equation}
\frac{4 \pi}{c} \int dE_\nu \, E_\nu \, J_{{\rm all} \, \nu} (E_\nu) \simeq \omega_\nu < \frac{3}{5} \omega_\gamma \, .
\end{equation}   
The most recent result from Fermi-LAT~\cite{Abdo:2010nz} translates
into an energy density of $\omega_{\rm tot} \simeq 5.8 \times
10^{-7}~{\rm
  eV/cm}^3$~\cite{Berezinsky:2010xa,Ahlers:2010fw}. Assuming an
$E_\nu^{-2}$ neutrino spectrum between energies $E_-$ and $E_+$ a
numerical simulation gives a cascade limit of~\cite{Markus}
\begin{equation}
E_\nu^2 J^{\rm cas}_{{\rm all} \nu} (E_\nu) \simeq \frac{3 \times 10^{-7}}{{\rm log}_{10}
(E_+/E_-)}~{\rm GeV} \, {\rm cm}^{-2} \, {\rm s}^{-1} \, {\rm sr}^{-1} \, .
\end{equation}

The diffuse neutrino flux has an additional component originating in
the energy losses of UHECRs {\em en route} to
Earth~\cite{Beresinsky:1969qj}. The accumulation of these neutrinos
over cosmological time is known as the cosmogenic neutrino flux.  The
GZK reaction chain generating cosmogenic neutrinos is well
known~\cite{Stecker:1978ah}. However, the normalization of the
neutrino flux depends critically on the cosmological evolution of the
CR sources and on their proton injection
spectra~\cite{Yoshida:pt,Engel:2001hd}. It also depends on the assumed
spatial distribution of sources; for example, local sources in the
Virgo cluster~\cite{Hill:1985mk}, would dominate the high energy tail
of the proton spectrum.  Another source of uncertainty is the energy
at which there is a transition from Galactic to extragalactic CRs as
inferred from a change in the spectral slope~\cite{Fodor:2003ph}.  A
fourth source of uncertainty is the chemical composition of the parent
CRs -- if these are heavy nuclei rather than protons, then the
neutrino flux is
reduced~\cite{Hooper:2004jc,Ave:2004uj,Allard:2006mv}.  The most
up-to-date calculations of the cosmogenic neutrino flux combine a
double-fit analysis of the energy and elongation rate measurements to
constrain the spectrum and chemical composition of UHECRs at their
sources; injection models with a wide range of chemical compositions
are found to be consistent with
observations~\cite{Anchordoqui:2007fi,Kotera:2010yn}. 

\begin{figure*}[t]
\begin{center}
\includegraphics[width=0.32\linewidth]{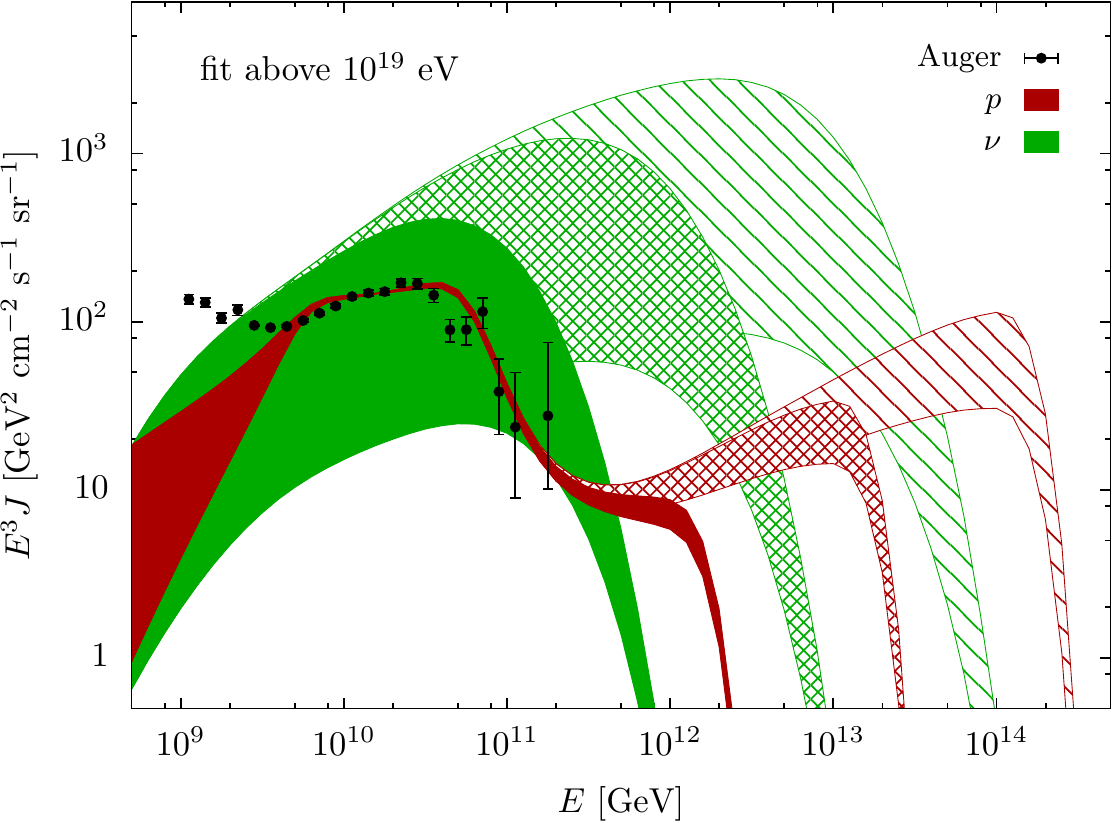}
\includegraphics[width=0.32\linewidth]{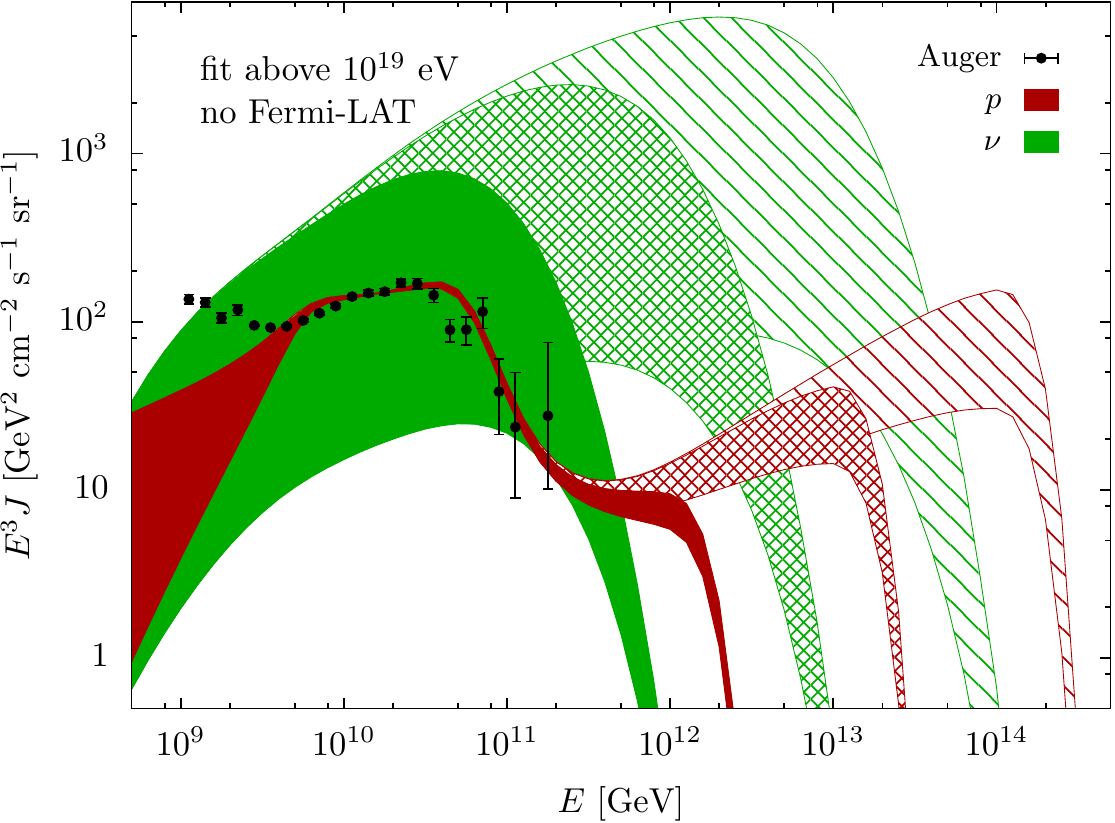}
\includegraphics[width=0.32\linewidth]{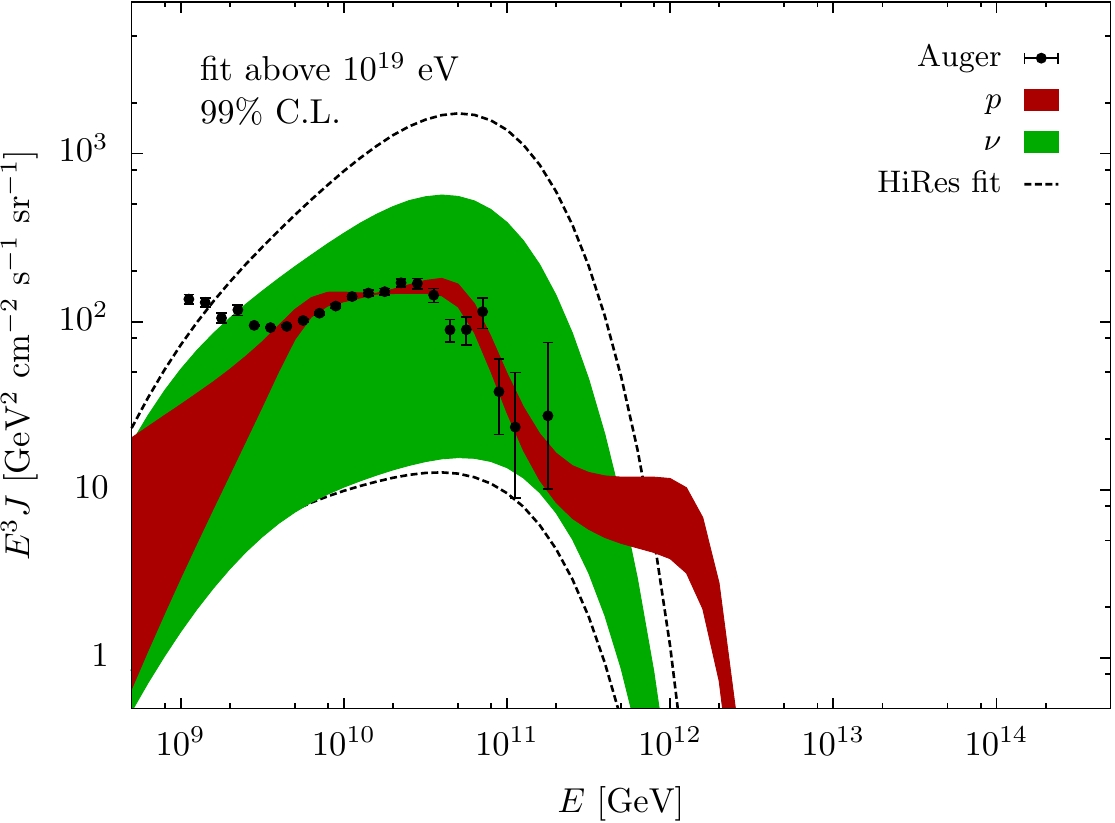}
\end{center}
\vspace{-0.3cm}
\caption[]{ Optimistic flux of cosmogenic neutrinos consistent with
  Auger data. The proton and GZK neutrino spectra from the 95\%
  C.L.~of the GOF test assuming $E_{\rm max}$ at $10^{21}$, $10^{22}$
  and $10^{23}$~eV are shown in the left and middle panels.  The left
  panel shows neutrino fluxes also consistent with the Fermi-LAT
  constraint. This constraint is relaxed on the fluxes shown in the
  middle panel. The right panel shows the proton and GZK neutrino
  spectra from the 99\% C.L.~of the GOF test assuming $E_{\rm
    max}=10^{21} eV$. The dashed lines show the results
  of~\cite{Ahlers:2010fw} for the GZK neutrino flux from a fit to the
  HiRes data with the same model parameters~\cite{GAPnote}.}
\label{gzk-nu-fluxes}
\end{figure*}

\begin{figure}[t]
\begin{center}
\includegraphics[width=0.49\linewidth]{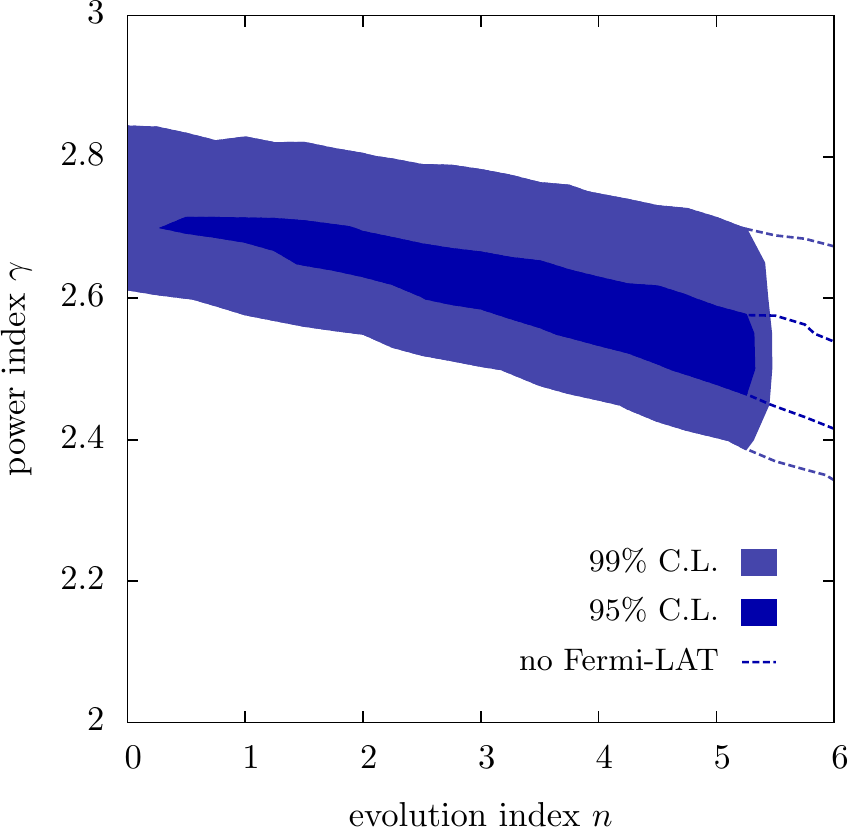}
\includegraphics[width=0.49\linewidth]{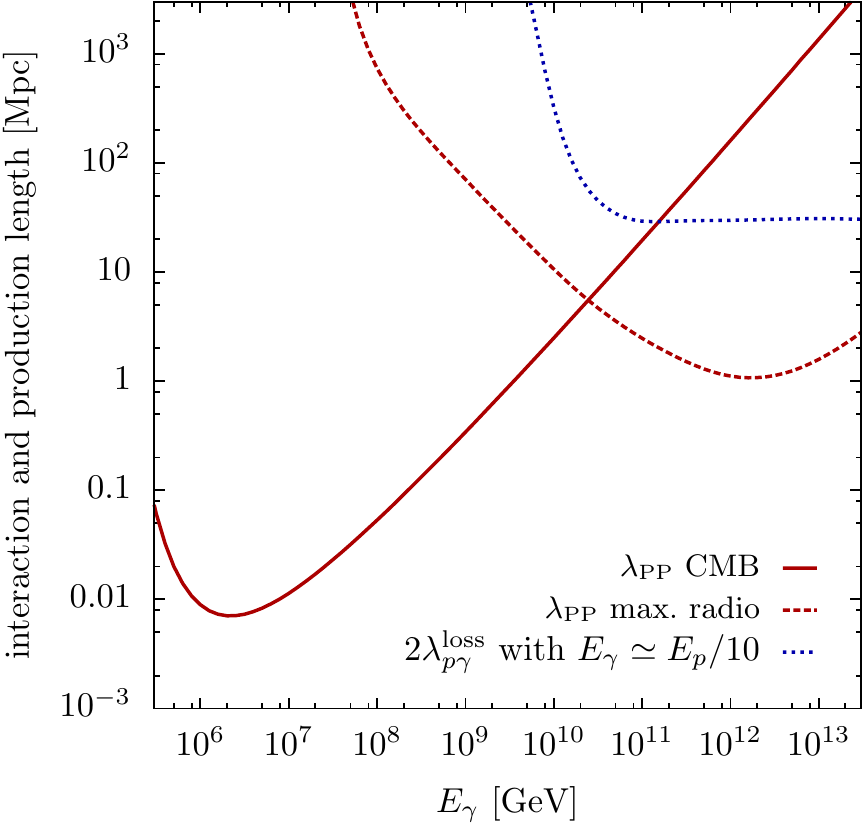}
\end{center}
\caption[]{{\it Left panel:} Goodness of fit test of the Auger data
  above $10^{19}$~eV. The dashed lines show the result without
  imposing the Fermi-LAT cut on the energy density. {\it Right panel:}
  Comparison of the $\gamma$-ray production length on the CMB
  (approximated as twice the total energy loss length in $p\gamma$
  interaction with $E_\gamma\simeq E_p/10$) with the PP length of
  $\gamma$-rays in the CMB and radio
  background~\cite{Protheroe:1996si}. The latter is always smaller
  than the production length and hence the contribution of
  $\gamma$-rays at high energies is suppressed~\cite{GAPnote}.}
\label{gof-intlen}
\end{figure}

Very recently, duplicating the goodness-of-fit (GOF) test from
Ref.~\cite{Ahlers:2010fw} optimistic fluxes of cosmogenic neutrinos
have been derived through normalization to the Auger
spectrum~\cite{GAPnote}. These optmistic fluxes, shown in
Fig.~\ref{gzk-nu-fluxes}, have been presented for a generous range of
maximal emission energies $10^{21}~{\rm eV} < E_{\rm max} <
10^{23}~{\rm eV}$.\footnote{By pushing the limit of what it is
  understood about astrophysical accelerators it is conceivable that
  $E_{\rm max}$ could extend as high as $10^{23}~{\rm eV}$.}  The GOF
test of the compatibility of a given model (characterized by the
injection spectral index $\gamma$ and cosmic evolution index $n$) with
the Auger data is shown in Fig.~\ref{gof-intlen}. The GOF test with
Auger data requires a sharp cutoff at the source on the minimal
emission energy of CRs, $E_{\rm min}=10^{19}~{\rm eV}$.  Such a
particular threshold can be explained if, e.g., the sources are
optically thin and the ambient optical photon plasma is significantly
suppressed, so that  $\tau_{\rm int}(E<E_{\rm min} ) \gg \tau_{\rm
  int}(E>E_{\rm min})$.

Considering the large neutrino fluxes shown in
Fig.~\ref{gzk-nu-fluxes} one might worry about a strong contribution
of UHE $\gamma$-rays from pion production and a violation of limits on
the photon fraction set by Auger~\cite{Abraham:2009qb}. However, this
contribution is expected to be dramatically reduced due to PP on the
CMB and radio background. Figure~\ref{gof-intlen} shows a comparison
of the pair production length on the CMB and the (maximal) radio
background of Ref.~\cite{Protheroe:1996si} with the production length
of $\gamma$-rays via $p\gamma$ interactions on the CMB. (The latter
has been estimated as twice the production length of pions, assuming
that $E_\gamma \simeq E_p/10$.) The PP interaction length is always
much smaller than the production length. Hence, only very local
$\gamma$-ray sources can contribute with a small fraction to the
overall photon fraction (see e.g.~Ref.~\cite{Ahlers:2011sd}).

Assuming an observational efficiency of 25\%, the JEM-EUSO sensitivity
 to EEC$\nu$'s has been computed considering both nadir and tilted
sky-surveys. As shown in Fig.~\ref{nus}, after 3~yr of running the
instrument will be sensitive to neutrino fluxes below the WB and
cascade limits. Thus, in principle, if there exists powerful
astrophysical sources that are capable of accelerating protons up to
$E \sim 5 \times 10^{21}~{\rm eV}$, with a power law spetrum $\propto
E^{-2}$, JEM-EUSO will be able to observe secondary neutrinos
originating in the interactions of these protons both at the source
and on their way to Earth.\footnote{Recall that $E_\nu \sim E/20$.} If
we take a more conservative view, in which we try to reproduce the
UHECR spectrum measured on Earth and Fermi-LAT observations, the
expected EEC$\nu$ flux yields a prediction just at the threshold of
JEM-EUSO observational capacity, see Figs.~\ref{nus} and~\ref{gzk-nu-fluxes}.

On the other hand, the ability to study neutrino interactions at such
extreme energies will open a unique window on possible physics beyond
the Standard Model of strong and electroweak interactions. While
astrophysical sources (at present level knowledge) can accelerate
particles to energies $10^{21}~{\rm eV}$ at most,
topological defects can produce particles, including neutrinos, up to
the GUT and Planck scales. Therefore, detection of neutrinos $E_\nu \agt
10^{21}~{\rm eV}$ would be a direct signature of  physics at energies well beyond the reach of laboratory experiments, such as signatures of  topological defects, e.g., the recently proposed~\cite{Berezinsky:2011cp} string-cusped model.

\section{Global Light System (GLS)}
\label{sec:GLS}
\subsection{Overview}

An accurate reconstruction of cosmic ray observables, such as the energy and direction of each event, the shower profile, and the number of events within a given energy, are key for identifying the sources of the highest particles in the universe. To accurately reconstruct these observables, three functions of the pioneering JEM-EUSO instrument are critical:
\begin{enumerate}
\item Triggering: How efficiently JEM-EUSO triggers on the optical
  signatures of EAS and discriminates against background signals.
\item Intrinsic luminosity measurement: How accurately JEM-EUSO measures the
  intrinsic luminosity of the EAS.  Intrinsic luminosity is determined
  by applying properly the corrections for geometric, optical, timing,
  and atmospheric effects.
\item Pointing Accuracy: How accurately JEM-EUSO measures the
  arrival direction of the EAS over the entire sky.
\end{enumerate}

The JEM-EUSO trigger must become efficient for EAS energies above $3 \times 10^{19}~{\rm eV}$. At the same time, the false trigger rate must be low enough for the onboard trigger processing system. To accomplish this, the trigger needs to be tuned carefully for optimum performance under varying atmospheric conditions and background light levels.

The energy of each cosmic ray is determined from the measured intrinsic luminosity of the EAS. Measurements made by JEM-EUSO must be corrected for scattering light in the intervening atmosphere between the ISS and the EAS. The processes for correcting these luminosity measurements and identifying atmospheric regions of the FOV that cannot be used, must be tested in a variety of atmospheric conditions and tuned to minimize the error.

Measurements of each EAS made by JEM-EUSO must be used to reconstruct the EAS trajectory in the atmosphere and used to determine the arrival direction of the cosmic ray in celestial coordinates. This requires not only observing the EAS over a sufficient length in the atmosphere but also correcting for distortion in the optics and pointing errors that result from the flexing of the ISS structure. The pointing accuracy of JEM-EUSO must be checked and the methods for arrival direction reconstruction optimized. Also required are in situ characterizations of the measurement accuracy as a function of
time. Factors that vary with time include, for example, atmospheric conditions, background ground light levels, the tilt angle of the detector, and the age of the instrument as it races over the Earth at
$28,000~{\rm km/h}$ in this pioneering mission.  

An onboard monitoring
system of UV light-emitting diodes (LEDs) is operated during daylight
segments of the flight when the aperture door is closed. This
monitoring function measures any relative variation of individual
multi-anode photomultiplier tube (MAPMT) performance and transmission
through the lenses during the mission. The atmosphere conditions in
the vicinity of an EAS are characterized using JEM-EUSO to measured
back-scattered light from an onboard steerable laser.  Together these
function as a Light Detection And Ranging (LIDAR) system. An IR
camera, combined with measurements by JEM-EUSO operating in a slow
data acquisition mode, characterize the background photon level,
altitude and opacity of any low-lying clouds present. This data,
together with a model of the atmosphere and the location and
inclination of the EAS, are used to correct for absorption and
scattering losses suffered by the fluorescence and Cherenkov photons
in each EAS.

The proposed GLS is a worldwide network
that combines ground-based Xenon flash lamps and steered
UV lasers to provide the most practical and cost-effective alternative to
validate the three key parameters (see Table~I). The GLS will generate benchmark
optical signatures in the atmosphere with similar characteristics to
the optical signals of cosmic ray EAS. But unlike air showers, the
number, energy, precise time, direction (lasers) can be specified. The
local measurement by monitors in the GLS units and the remote measurements
made by JEM-EUSO can then be compared.  Specifically,  JEM-EUSO will
reconstruct the pointing directions of the lasers and the  intensity of
the lasers and flash lamps to monitor the detector's triggers, and
accuracy of energy and direction reconstruction. The GLS provides the
means to tune the analysis for these atmospheric effects, verify
performance of the instrument subsystems (optics, focal surface
detector calibration, LIDAR and IR camera, attitude determination) and
validate the software and models used in event reconstruction for
these space-based observations of EAS.\footnote{There are no identified 
UHECR standard candles to provide
the convenient references, such as test beams of $10^{20}~{\rm eV}$ particles. 
The proposed GLS will provide the best alternative to a test beam.}

\begin{table}
\label{table1}
\caption{The GLS will include three types of units.}
\begin{tabular}{c|c|c}
\hline
\hline
~~~GLS unit~~~ & Sources & ~~~\#~~~ \\
\hline
GLS-X & Xenon Flashlamps (XF) & ~~~6~~~ \\
GLS-XL & XF and Laser & ~~~6~~~ \\
~~~GLS-AXHL~~~ & ~~~XF and Horizontal Laser~~~ & ~~~1~~~ \\
 & (airborne) & \\
\hline
\hline
\end{tabular}
\end{table}

There will be 12 ground based units strategically placed at sites
around the world. Six locations will have flashers and a steerable and remotely operated
laser (GLS-XL)  and 6 will have flashers only (GLS-X). Sites will be
chosen for their low back-ground light and altitude (above the
planetary boundary layer). The placement of GLS units will cover a
wide range of terrestrial conditions (land, open ocean) geographical
locations, latitudes, and altitudes. It will test the full range of
cloud conditions and track trajectories that are encountered in the
mission. A nonexclusive list of candidate sites is provided in  Table~II. We envision
placing the 6  GLS-XL units at sites with available grid power, although
solar power (as used at Auger)  will be considered if the scientific and 
logistical justification is exceptional and unique. The 12
units will be supplemented by campaign style measurements with an
airborne system, that features a Xe flashlamp and a horizontal UV
laser. Dubbed the GLS-AXHL, it will be mounted in in P3B Orion
aircraft stationed at Wallops Flight Facility/NASA (WFF) in Virgina USA. The
GLS-AXHL will be flown over the open ocean at selected altitudes under
JEM-EUSO. Details are given in Sec.~\ref{sec:operations}.

\begin{table}
\label{table2}
\caption{Candidate locations for deploying GLS.}
\begin{tabular}{c|c|c}
\hline
\hline
Location & ~~~Latitude~~~ & ~~~Altitude~~~ \\
\hline 
Jungfraujoch (Switzerland) & $47^\circ$~N & 3.9~km \\
Mt. Washington (NH, USA) & $44^\circ$~N & 1.9~km \\
Alma-Ata (Kazakhstan) & $43^\circ$~N & 3.0~km \\
Climax (CO, USA) & $39^\circ$~N & 3.5~km \\
Frisco Peak (UT, USA) &   $38^\circ$~N  & 2.9~km \\
Mt Norikura (Japan) & $30^\circ$~N & 4.3~km \\
Mauna Kea (HI, USA): & $20^\circ$~N & $> 3.0$~km \\
Nevado de Toluca (Mexico) & $19^\circ$~N & 3.4~km \\
Chacaltaya (Bolivia) & $16^\circ$~S & 5.3~km \\
La Reunion (Madagascar)  &$21^\circ$~S & 1.0~km \\
Cerro Tololo (Chile) & $30^\circ$~S & 2.2~km \\
Sutherland (South Africa) & $32^\circ$~S & 1.8~km \\ 
Pampa Amarilla (Argentina) & $35^\circ$~S & 1.4~km\\ 
South Island (New Zealand) & $43^\circ$~S & 1.0~km \\
\hline
\hline
\end{tabular}
\end{table}

In one orbit JEM-EUSO views an area of $1.6 \times 10^7~{\rm
  km}^2$. Each day JEM-EUSO sweeps over an area of $2.5 \times
10^8~{\rm km}^2$  (which is 83\% of Earth's area between $\pm 51.6^\circ$
latitude). On average a GLS site will be
over-flown with favorable atmospheric conditions (assuming 10\% duty
cycle for a ground based station) each day of the mission providing a
data base that will be sensitive to seasonal variations at the
locations. During an over-flight, the GLS unit will produce calibrated
signals repeatedly while in the JEM-EUSO FOV, which last 56 seconds on
average. 

Our plan addresses the several challenges presented by the GLS.
 The GLS is developed using Commercial Off-The-Shelf (COTS)
components, assembled and tested at the University of Alabama
Huntsville (UAH), Colorado School of Mines (Mines) and the Marshall
Space Flight Center (MSFC). The fabrication, testing, operations,
analysis and maintenance of these units begin only after the mission
has been confirmed by the Japan Aerospace Exploration Agency (JAXA)
(estimated to occur in 2014). Prior to confirmation by JAXA, the GLS
designs will be developed, prototyped and tested. A GLS prototype will
support a high-altitude balloon flight in  2014 of a prototype
JEM-EUSO telescope, named EUSO-Balloon. The Centre National d'Etudes Spatiales (CNES) is the balloon mission sponsor and this GLS prototype is a key element of
this sub-orbital program that will demonstrate the readiness of the
technology used for the spaceflight mission. The experience gained
during the sub-orbital flight will assure that the GLS design 
and operations can meet the requirements of the JEM-EUSO 
spaceflight mission.

\subsection{The Xe-flasher (GLS-X)}

\begin{figure}[t]
\begin{center}
\includegraphics[width=0.42\textwidth]{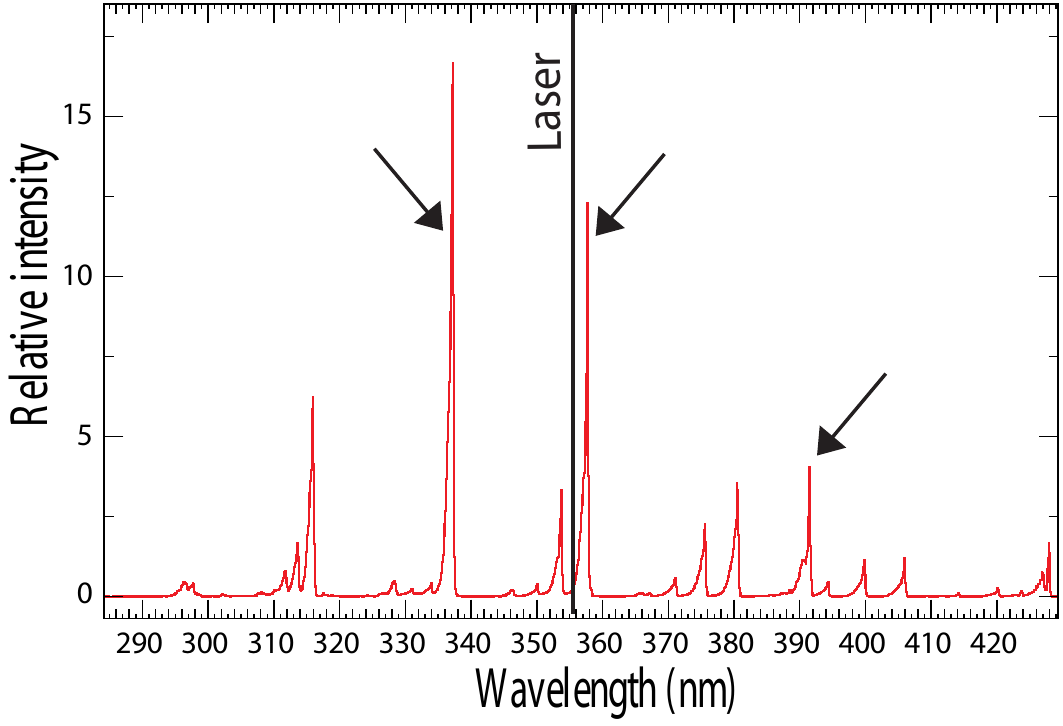}
\caption{Measured fluorescence spectrum~\cite{Ave:2007xh}. The
  wavelength of the GLS flashers will be filtered to produce light at
  the lines indicated. The wavelength of the GLS lasers is also shown,
  (adapted from Ref.~\cite{Ave:2007xh}.)}
\label{Lawrence1}
\end{center}
\end{figure}

Since JEM-EUSO will view the Earth's surface, intrinsic luminosity can
be monitored directly with flash lamps. All GLS stations will include
four flashlamps, filtered to match the three primary UV fluorescence
yield
lines~\cite{Arqueros:2008cx,Arqueros:2008en,Ave:2007em,Ave:2007xh} at
391~nm, 357~nm, and 337~nm (see Fig.~\ref{Lawrence1}) and a Schott BG3
wide band-pass filter, identical to that covering the focal plane
detector.

\begin{figure}[t]
\begin{center}
\includegraphics[width=0.42\textwidth]{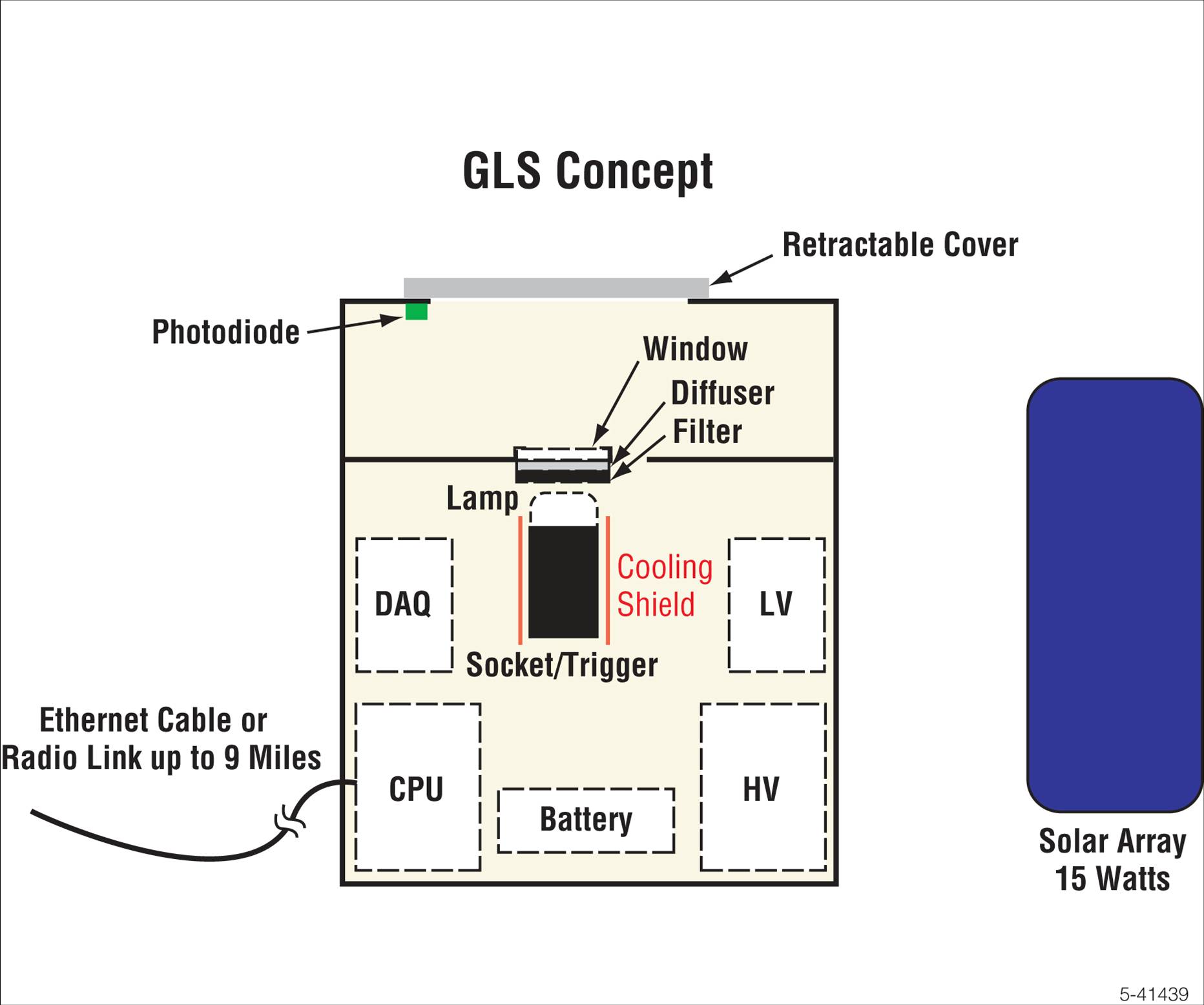}
\caption{Diagram of a single GLS-X Concept showing the basic elements.}
\label{Marc1}
\end{center}
\end{figure}

\begin{figure*}[t]
\begin{center}
\includegraphics[width=0.49\linewidth]{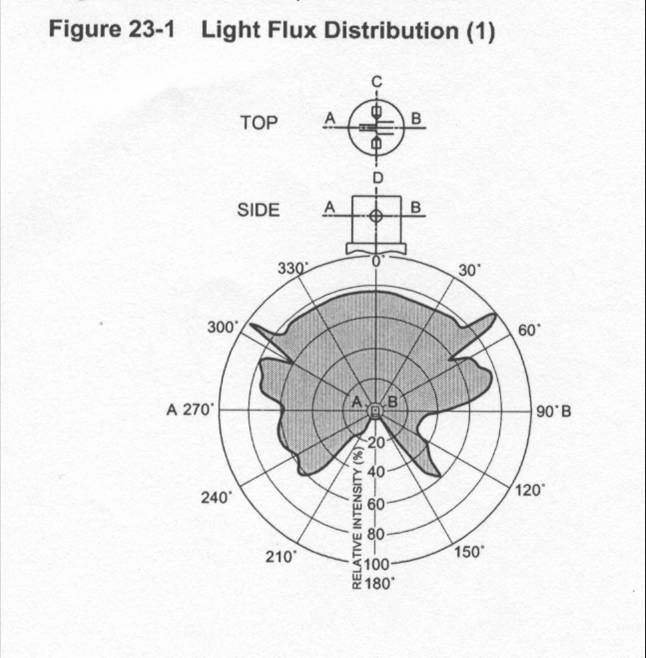}
\includegraphics[width=0.49\linewidth]{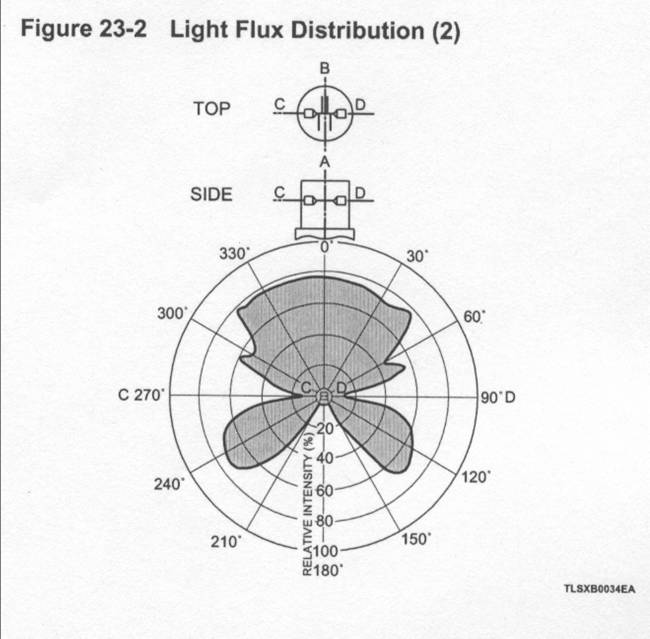}
\end{center}
\vspace{-0.3cm}
\caption[]{Projections of the relative intensity output for the flash lamp. The FOV for JEM-EUSO is restricted to ±30° from the vertical where the output is smooth. }
\label{Marc34}
\end{figure*}

The critical functional requirement is a high reliability
light flash. Our GLS-X concept incorporates four individual
UV-flashers (Hamamatsu L6604) that serve as a set of “standard
candles”. Each flasher is fitted with a mechanical aperture, clear
glass outer window, followed by a glass diffuser and one of the
optical filters (see Fig.~\ref{Marc1}). Each flasher includes other
ancillary components from Hamamatsu to meet the performance
specifications: trigger module, cooling jacket, capacitors for the
initial and primary flashes and a power supply. The L6604 has a highly
stable output, varying $<3\%$ from flash-to-flash, long stable
lifetime with more than $10^7$ flashes and $<3\%$ degradation over the
lifetime of the mission~\cite{Hamamatsu}. The light pattern from each
flash is smoothly distributed over a wide field (see
Fig.~\ref{Marc34}). The intensity of the flash varies slowly over a
field $>60^\circ$ without any spikes or valleys in the intensity. The
ripples in the light intensity at the outer fringes of the light field
fall outside of the aperture and will not be visible to
JEM-EUSO. These key performance parameters have been verified in
laboratory tests.

\begin{figure}[t]
\begin{center}
\includegraphics[width=0.45\textwidth]{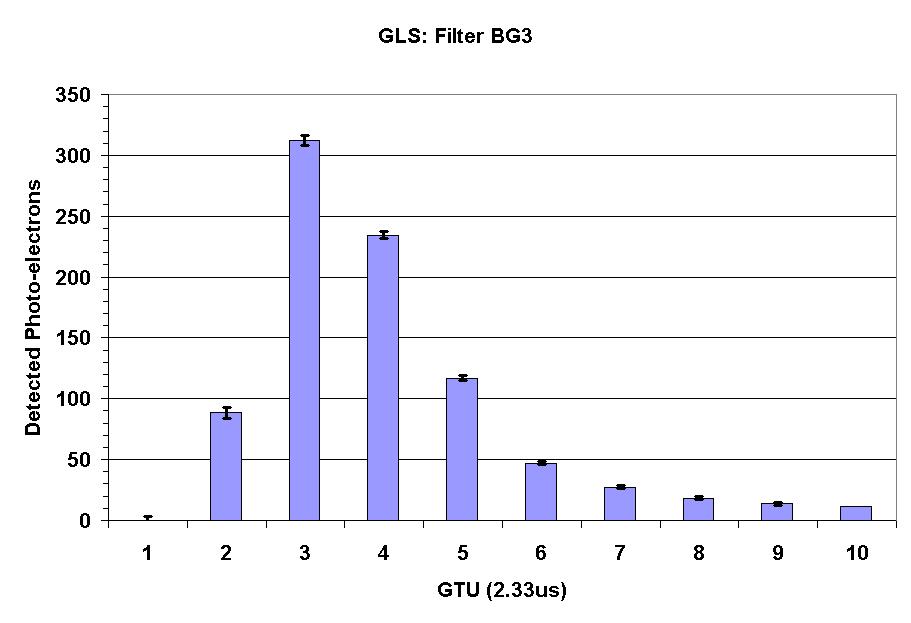}
\caption{Measured average  
   intensity of a L6604 flash lamp with a pulsed width
  stretched to more than 10~$\mu$s. The error bars indicate the
  standard deviation measured for 100 pulses. The vertical axis is
  scaled to show the estimated number of photo-electrons detected by
  JEM-EUSO during 
  each of the 10 GTUs shown.
  }
\label{Marc5}
\end{center}
\end{figure}

To better match the timing characteristics of a
vertical EAS as seen by JEM-EUSO, the light pulse are stretched to
$>10$ microseconds. This is the minimum acceptable duration for the
on-board trigger algorithm and enables JEM-EUSO to detect the GLS-X
flashes without requiring a separate dedicated trigger mode. The
capacitance and voltage are the principal parameters that control the
timing and intensity characteristics of the flashes. We have
investigated the performance of the flash lamps for various capacitor
values and have succeeded in stretching the pulse while maintaining
the stability and repeatability of the light flashes. We verified the
performance in the laboratory by analyzing 100 flashes shown in
Fig.~\ref{Marc5}. The data has been converted to photo-electrons
detected by the JEM-EUSO instrument based on estimated loses for a
clear nighttime atmosphere and the instrument's
efficiency factors. The standard deviation for the  100 pulses,
measured using the 2.33~$\mu$s
GTU employed in JEM- EUSO, is significantly smaller than the variation
expected due to photoelectron statistics for the signal recorded on
board; typically 30\% smaller. The data in each pixel for successive
GTUs are summed together to determine the total signal detected from
each GLS-X flash.  

The GLS-X flashers are housed in a weather tight
enclosure that includes the flasher and its ancillary components, a
photo-diode and data acquisition system (DAQ), a mechanical shutter
and motor, a single board computer (SBC), solar array and battery (if
needed) and a remote communication element (RCE). The RCE provides
communication capability with the Central Operations Center (COC)
located at the National Space Science and Technology Center (NSSTC) to
upload and download parameters for the GLS-X and to retrieve data to
determine its operational status. The RCE will depend on the on-site
resources available and will use either a satellite modem or wireless
router. The SBC controls the operations of the GLS-X: performs
housekeeping functions and reports status to the COC. The GLS-X units
are mounted on top of steel poles to minimize potential interferences
from the immediate local environment. Discussions with personnel in
the US forestry service experienced with remote sensing operations
have been used to develop our initial plans for the sites. Members of
the international JEM-EUSO Collaboration will provide assistance with
identifying candidate sites for the GLS-X and assistance in
identifying and communicating with local authorities to assure the
operation is free from any interference prior to selecting a
site. Table~II lists several candidate sites that meet the key
requirements for the GLS-X: high elevation and low light pollution.

\subsection{The GLS-XL} 

\begin{figure}[t]

\begin{center}
\includegraphics[width=0.42\textwidth]{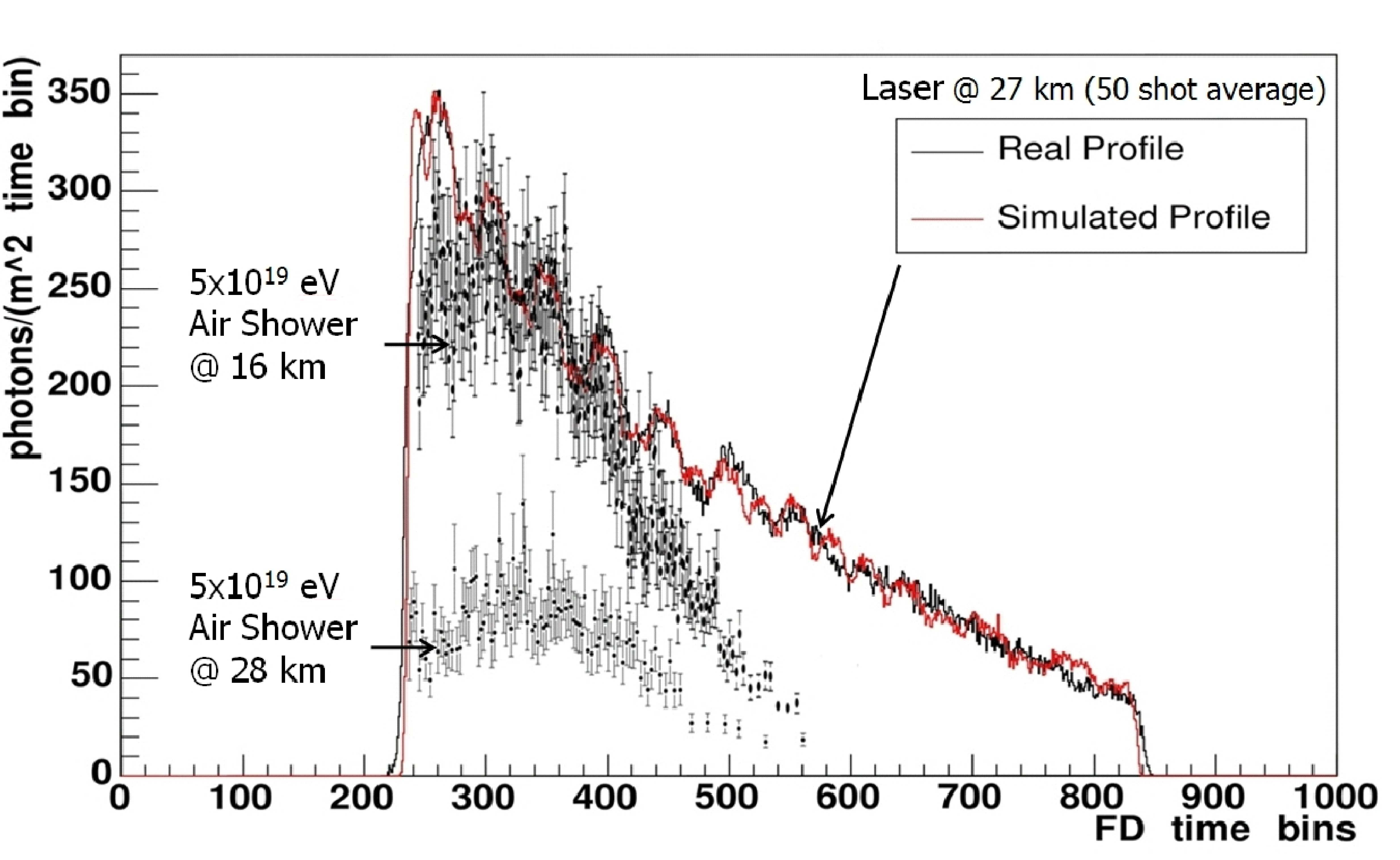}
\caption{Longitudinal profiles of vertical laser shots and near-vertical cosmic ay air showers recorded by the Pierre Auger fluorescence detector from a horizontal distance of 27~km \cite{Wiencke:2008cj}.  One time bin is 100~ns. The viewed laser track is 12~km long in this example.  The downward going EAS profiles have been flipped so that the left edge of all profiles corresponds to the bottom of the FD field of view.}
\label{Lawrence2}
\end{center}
\end{figure}

Six GLS units will have steered UV laser systems that leverage the
successful experiences of the pioneering Fly's
Eye~\cite{Baltrusaitis:1985mx,Bird:1994uy},
HiRes~\cite{Wiencke:1999pi,Lawrence2r,Cannon:2003wf}, and the Pierre Auger
Observatory (Auger)~\cite{Wiencke:2011yc,Abreu:2011pg}. Benchmark data
from automated remote lasers~\cite{Arqueros:2005yn,Mines} in the Auger
fluorescence detector (FD)~\cite{Abraham:2009pm} FOV have proven critical both for
measuring the performance of the FD~\cite{Abraham:2010pf,:2010zzl} and for realizing the
science program~\cite{Abraham:2010yv,PierreAuger:2011aa}. There is an approximate effective optical
equivalence between a 5~mJ 355~nm laser track from pulse shot crossing
the field of view, and a $10^{20}~{\rm eV}$ air shower
track~\cite{Wiencke:2006hg,Wiencke:2008cj} (see
Fig.~\ref{Lawrence2}). 

\begin{table}
\label{table3}
\caption{GLS-XL laser parameters.}
\begin{tabular}{c|c}
\hline
\hline
Parameter & ~~~Specifications~~~ \\
\hline
wavelength & 355~nm \\
energy/pulse & 1 - 10 mJ\\
pulse width & 10~ns\\
energy calibration (relative) & 3\%\\
~~~energy calibration (absolute)~~~ & $<$10\% \\
pointing (relative) & $0.02^\circ$ \\
pointing (absolute) & $0.2^\circ$\\
timing (absolute) & 50~ns \\
\hline
\hline
\end{tabular}
\end{table}

A similar equivalence will also apply to JEM-EUSO.
In this case both profiles will be elongated in upper
part of the troposphere where the air density is less. By adjusting the energy of the
laser, the triggering threshold for track-like optical signatures will
also be tested. Lasers can also be aimed at UHECR potential sources over the full sky (north and south).
A selection of ~20 sources could include, for example, the galactic
center, Cen A, Virgo, and other
objects of astrophysical interest.  A sky map of reconstructed laser track directions will be accumulated over the mission.
The clusters of points and their spread about the directions of the targeted sources,
 will provide a simple but comprehensive validation of the absolute EAS pointing
accuracy reconstruction by JEM-EUSO. Included in this test is transfer of time stamps from a precise clock on ISS
through the JEM-EUSO hardware and the onboard and ground based data analysis chains. 
During clear periods when 355~nm light propagation in the atmosphere between the GLS site and the ISS can be 
described to the few percent level using molecular scattering alone (i.e. aerosol optical
depth $\ll$  molecular optical depth),  it will be possible to test intrinsic
luminosity with both the direct Xe flasher light (point source) and the
scattered UV laser light (track source).  These periods can be identified
by comparing the ratio of flasher measurements at different viewing
angles by JEM-EUSO to the ratio predicted by molecular scattering together with measurements of the 
JEM-EUSO IR camera and onboard laser. 

The laser subsystem of the GLS-XL will use
many of the components and design features used at the Pierre Auger
laser facilities. Proposed specifications are listed in Table~III. The
laser will be a frequency tripled YAG. Proven models at HiRes and
Auger include BigSky (Quantel) CFR and Centurion \cite{Quantel}. The optics will
include harmonic separator mirrors to remove residual primary and
secondary harmonics. The net polarization of the beam will be
randomized so that the atmosphere scatters the same amount of light
azimuthally about the beam direction. The steering mechanism will use
COTS controller and two orthogonal rotational stages following a
design used by Auger and HiRes.  

The steering mirrors will be coated for 355~nm and 632~nm reflections so that they
can be aligned using a HeNe laser device with an internal pendulum level. These are
available at building supply stores. The relative energy of each laser pulse will be measured by
 directing a small fraction of the
beam into a pyroelectric energy probe. The absolute energy of the beam
will be calibrated when steering mechanism is parked directing the
beam on to a second energy probe mounted inside the system
enclosure. To facilitate identification of laser data, the laser will
be triggered at precise times referenced to the global positioning system (GPS)  timing
 and the times of each shot recorded locally.
Since the laser will be far from the JEM-EUSO detector, the beam
will be much narrower than the area viewed by a pixel. Consequently,
the beam can be de-collimated significantly to reduce the eye-safe distance.

The system will be housed in a commercially available temperature controlled industrial
enclosure/cabinet designed for outdoor use. The steering head will be
mounted on top. The laser, optics and controls and other instrumentation will be mounted
inside. The entire system will be computer controlled using a single
board computer (SBC) with no moving parts. The SBC will be connected to
the internet through a firewall computer. Depending on the site a microwave
wireless link will be used.  (30 km links have been used succesfully for many
years at the Auger laser systems.) The system will be operated and reprogrammed remotely.  
The system will include wind, rain, and temperature sensors and a
webcam. Past experience has shown that these laser systems are the
most stable, and thus provide the most useful benchmark data, when
their temperature is controlled ($20 \pm 5~^\circ{\rm C}$) and the optics are kept
free of dust. This compact arrangement, making use of a COTS enclosure
is preferred after long experience with other structures.
Consequently, it will be necessary to design and build a  GLS-XL
prototype and test an entire unit extensively, including the
installation and alignment procedures, before finalizing the design
and beginning fabrication of the 6 GLS-XL systems planned for the
JEM-EUSO ISS mission.

\subsection{Mission Operations}
\label{sec:operations}

The GLS-to-EUSO interface consists of a trigger pattern that includes
the range of intensities and timing characteristics for the
GLS. During nominal mission operations, observation planning will be
routinely conducted. The resulting plans will be uploaded every few
days to the ISS and transferred to EUSO, and executed
autonomously. The GLS flashes and tracks will be processed in the same way as
EAS triggered events and included in the telemetry stream downloaded
to the JAXA operation center and transferred to the JEM-EUSO science
center. This data also includes data generated by the onboard
Atmospheric Monitoring System (AMS) and the instrument housekeeping data.

The ISS will overfly a GLS-site with favorable weather conditions
every  night, on average. The ISS ground track speed is $7~{\rm
  km/s}$ and an individual pixel in the focal detector plane has a
footprint on the ground of $460 \times 460~{\rm m}^2$ near the nadir
for an ISS altitude of 350~km. A GLS site will be in the JEM-EUSO FOV
an average of 56 seconds; a pixel crossing time lasts for 60
milliseconds and an average of 1300 pixels will view the GLS site. The
GLS-X operates at 10~Hz.  The GLS-XLs will alternate laser and
flasher pulses, for a total rate of 20~Hz to provide nearly a
continuous set of measurements across the JEM-EUSO FOV. For each
trigger, the onboard AMS  is automatically activated and acquires
an IR camera image and a LIDAR shot at the location of the
GLS~\cite{Adams:2003cu}. The signal detected by EUSO under clear sky conditions
is estimated to
be $>800~{\rm p.e.}$/ xenon flash~\cite{Adams:2003cu}.  The
corresponding estimate for the lasers is $\sim 30~{\rm
  photons/m}^2$ (at aperture) /pixel/mJ of laser energy (low elevation
angle direction) for clear sky conditions.  This translates to $500-1000~{\rm p.e.}$
per track for 5 mJ laser pulses, depending on the pointing direction.
 Multiple flashes of the GLS
during each over-flight produce transmission estimates with an
accuracy of a few percent. The absolute UV attenuation will be
determined by analysis and compared with the atmospheric attenuation
based on the AMS. There will occasionally be very clear
conditions when the measured total optical depth is not significantly
greater the molecular optical depth. The latter can be determined
accurately~\cite{Abreu:2012zg} from the global data assimilation
system (GDAS)~\cite{NOAA}. In these cases, the intrinsic luminosity
resolution can be measured using track-like signatures by comparing
the laser energy as reconstructed by JEM-EUSO and as measured by the
GLS-XL energy probe.

The deployment of the GLS will occur prior to the launch in 2017. The
12 ground based sites (6 GLS-XL, 6 GLS-X) will be selected and the GLS
units will be installed and tested. Each site includes a
microprocessor or single board computer with internet access, enabling
remote management. Operations will be conducted remotely from the COC
located at the NSSTC in Huntsville. UAH/MSFC will be responsible for
programming the xenon flashers and monitoring their performance. The
Colorado School of Mines will be responsible for programming the
lasers and monitoring their performance. The ISS times and flash
intensity will be pre-loaded for each overpass, when predicted
atmospheric conditions are favorable. The pattern of laser shots will be programmed
 in advance and will include shots fired at potential
UHECR sources. After each over-flight, housekeeping data will be
relayed to the COC for analysis. Between ISS fly-overs, internal
diagnostics will be performed, typically once per day to make sure the
systems are functional.

The GLS-AXHL will be installed in a P3B airplane
managed by the NASA Airborne Science Program (ASP). The P3B 
features include an upward viewing portal that is
available to install a GLS-X and a side port that will be fitted with
a fused silica window to transmit the UV GLS-L pulses. Once the
necessary mechanical and electrical interfaces have been developed and
tested, the airplane will be deployed approximately monthly for under
flights of the ISS at night. The P3B will fly out 500~km from the
eastern seaboard to rendezvous with the ISS for a single under-flight.    
Each flight will target a
specific set of conditions. Since the earth rotates by 
15 degrees  every hour, there will be one ISS overpass 
per P3B flight. Over the length of the JEM-EUSO mission, these
flights will cover a range of altitudes, atmospheric and cloud
conditions, and moonlight. The flights will provide sufficient testing
for the large number of events acquired by JEM-EUSO over the
oceans. After each flight the GLS-AXHL will be dismounted and stored
at WFF until the next flight. The details of the P3B flight data will be
obtained and included in the analysis of the GLS-AXHL and JEM-EUSO
data. The capabilities of the P3B to support the spaceflight mission
have been evaluated by the ASP and they have agreed to provide the
support needed for these airborne operations.  

In addition to validating the accuracy of the EAS reconstruction
analysis, the GLS tests the focus of the optical system. The GLS-X
serves as a point source and the images of the GLS are small spots
with widths that are determined by the optics point-spread
function. Throughout the course of the mission, the GLS spot-size will
be analyzed to determine if the telescope is maintaining proper
focus. The downloaded images of the GLS will be analyzed and compared
with optic simulations to guide any focusing adjustments
required. Commands can be sent to the telescoping mechanism on
JEM-EUSO to adjust the distance between the focal plane and optics to
improve the focus.

\subsection{ EUSO-Balloon Flight Tests}
 
Balloon flights are planned to demonstrate a prototype of the
JEM-EUSO telescope. 
The EUSO-Balloon mission sponsored by CNES has the following key objectives:
\begin{itemize}
\item Full-scale end-to-end test of JEM-EUSO proof of concept and technique.
\item Operations of key electronic components including HV control and onboard trigger. 
\item Acceptance of signals over a large dynamic range.
\item Experimental determination of the effective UV background below 40 km.
\item Acquisition of JEM-EUSO type data.
\item Detection of laser induced events from space. 
\item Improve trigger algorithms with real data.
\item 1st imaging of EAS looking down on the Earth's atmosphere.
\end{itemize}

A prototype of the GLS-AXHL will be deployed to support the suborbital
EUSO-Balloon mission. In addition to providing real experience with supporting an
Earth observing EAS detector, the GLS-AXHL will provide critical data
for the above listed objectives. 

The GLS support for the balloon flights include a
ground-based laboratory measurement of the instrument response and an
under-flight using the P3B platform with a prototype of the airborne
GLS-AXHL planned for the spaceflight mission. Since the balloon will
 travel more slowly than the aircraft (unlike the ISS), 
the aircraft can fly multiple passes near the balloon to 
conduct multiple tests. These exposures will be
used to develop the GLS design and to exercise key GLS functions,
operations, and analysis for the spaceflight mission. A GLS prototype
will be taken to the instrument integration site in France. After
completing the integration of the EUSO-Balloon prototype telescope, ground
tests will be conducted to assess the overall performance of it. These
tests include exposures using LEDs, the dark night sky and bright
nighttime sources (moon and stars). The prototype GLS will be one of
the sources used in this preliminary performance assessment. In turn,
the tests can verify the GLS flash intensity and duration are
compatible with the onboard trigger algorithm and dynamic range of the
focal plane detector. Preliminary analysis of the acquired data will
be completed immediately and a more comprehensive analysis will be
completed after the instrument checkout period has been
completed. This preliminary exposure will be used to understand the
response of the telescope and improve the design of the GLS for the
balloon under flight to insure successful operations during the
balloon mission.  

The P3B platform will be prepared to accommodate the GLS-AXHL for
airborne operations. Well in advance of the balloon flight, the
integration and testing (I\&T) of the GLS-AXHL will be coordinated
with the ASP at NASA/WFF. The upward portal on the P3B will be used
for the GLS-X and the sideway viewing port for the GLS-HL. Viewing
windows will be installed in the P3B for the GLS-AXHL. A Schott BK7
glass window is baselined because its optical performance meets
requirements for the three principal nitrogen fluorescence lines (337,
357 and 391 nm). A quartz window will also be investigated. The I\&T
includes mechanical fixturing provided by ASP and power resources
already existing on the P3B. An engineering flight of the P3B,
followed by a 1 hour instrument checkout flight, will be completed at
WFF to insure proper installation of the sources. Pre-flight and
post-flight ground tests of the GLS- AXHL will be completed to insure
they are operational and meet the requirements.

A primary launch site for  EUSO-Balloon
flights is in Canada. The  flight and GLS-AXHL crews will be on
stand-by at WFF during the balloon launch window. Once the launch of
the  EUSO-Balloon instrument is imminent, the P3B will deploy
from WFF. The P3B will
arrive on site during the nighttime portion of the balloon flight. Once on
site and in communication with the ground personnel, the P3B will
execute oval flight patterns at preselected altitudes. One leg of the
flight path will take the P3B within 1 km of the balloon ground
tract where the GLS-X is operated. The other leg of the oval is ~10 km
distance from the balloon ground track and the laser will shoot
across the telescope FOV. The ground personnel will provide feedback
in near real-time on the detection of GLS signals to confirm the
data acquisition. This balloon mission provides a unique opportunity
to exercise and test several elements of the GLS-AXHL system and data
analysis for the spaceflight mission. It will be possible to investigate, for
example,  the impact of the atmosphere in a
single location over the range of altitudes spanned by the EAS.

The P3B is our primary option for supporting the spaceflight and
balloon flight activities. In the event the P3B will not be available
for the balloon launch window, the ASP has alternate aircraft
available to support the balloon under-flight at a lower total cost
for the planned operations. Additionally, there are several private
aviation companies in the vicinity of  these remote launch sites that can be
evaluated to support the balloon under-flight. There is no issue of
availability for the P3B for the spaceflight mission because ISS
under-flight opportunities are frequent and can be planned months in
advanced.

\section{Summary and Conclusions}

The JEM-EUSO mission will provide a breakthrough towards the
understanding of astrophysical and physical aspects of the universe at
extremely high energies. To deeply explore this new face of the cosmos
EECR facilities need to observe the full sky and must reach colossal
exposures, ${\cal O} (10^6~{\rm km}^2 \, {\rm sr} \, {\rm yr})$.
Space-based observatories will be a critical element in accomplishing
this endeavor. JEM-EUSO is the first step in space: a pioneer and
pathfinder in the field.

The high statistics and better experimental handles of the instrument
will enable us to locate the CR sources in the sky. The discovery of
the first clear source of CRs will be earth-shattering. After
more than 50 years, the nature of the strongest cosmic accelerators
will be revealed with fundamental consequences for particle
astrophysics, including neutrino and gamma ray astronomy. A side
benefit of finding nearby CR sources is that the same analysis
will provide information on the strength and coherence length of
extragalactic magnetic fields.

In addition, the measurement of the shape of the spectrum above
$E_{\rm GZK}$ will determine if the steepening observed in the CR
intensity is due to the effect of the GZK phenomenon or to the
limitations of CR accelerators ($E_{\rm max}$). To address this
ambiguity, JEM-EUSO's high statistics data set would allow us to
observe a spectral recovery above the GZK suppression if $E_{\rm max}
\gg E_{\rm GZK}$.  An optimist might even imagine the discovery of ZeV
neutrinos, the telltale signature of universe's topological
defects. All in all, we are confident that the pioneering observations
of the JEM-EUSO mission will open a new era in CR physics.

\section*{Acknoweledgements}

We thank Markus Ahlers, Paolo Privitera, and Michael Turner for
valuable discussions and comments. We are also grateful to Aimee Giles
and the Kavli Institute for Cosmological Physics at the University of
Chicago for the most enjoyable atmosphere during the Workshop which was funded by 
an endowment from the Kavli Foundation.

\end{document}